\documentclass[10pt]{article}

\usepackage{graphicx}
\usepackage{tabulary}
\usepackage{multirow}
\usepackage[margin=1in]{geometry}
\usepackage{titlesec}
\usepackage{subcaption}
\usepackage[labelfont=bf,textfont=it,font=small]{caption}
\usepackage{parskip}
\usepackage[comma,sort&compress,numbers]{natbib}
\usepackage{enumitem}
\usepackage{bibunits}
\usepackage{setspace}

\usepackage{amssymb}
\usepackage{amsthm}
\usepackage{amsmath}
\usepackage[ruled]{algorithm2e}
\usepackage{array}
\usepackage{todonotes}
\usepackage{url}
\usepackage{nicefrac}        
\usepackage{siunitx}
\usepackage{tikz,pgfplots}
\usetikzlibrary{decorations,
                arrows,
                shapes,
                trees,
                shapes.geometric,
                calc,
                arrows.meta,
                positioning,
                fit, 
                shapes.multipart,
                backgrounds}
\usepackage{fontawesome5}
\pgfplotsset{compat=1.16}
\usepackage{adjustbox}

\usepackage[T1]{fontenc}
\usepackage{times}
\usepackage{helvet}
\usepackage[pagewise]{lineno} 
\usepackage{booktabs}
\graphicspath{{Figures/}}
\DeclareGraphicsExtensions{.pdf,.png,.jpg}

\titlespacing\section{0pt}{6pt plus 1pt minus 0pt}{6pt plus 1pt minus 0pt}
\titlespacing\subsection{0pt}{6pt plus 1pt minus 0pt}{6pt plus 1pt minus 0pt}
\titlespacing\subsubsection{0pt}{6pt plus 1pt minus 1pt}{6pt plus 1pt minus 1pt}
\titleformat{\section}{\large\bfseries\sffamily}{\thesection}{1em}{}
\titleformat{\subsection}{\normalsize\bfseries\sffamily}{\thesubsection}{1em}{}
\titleformat{\subsubsection}{\normalsize\bfseries\sffamily}{\thesubsubsection}{1em}{}

\usepackage{hyperref}
\usepackage{bookmark}
\hypersetup{
    colorlinks=true,
    linkcolor={magenta},
    citecolor={blue},
    urlcolor={blue!80!black},
    breaklinks=true,
	plainpages=true
}

\newcommand{\GRATEVtwo}{\texttt{GRATEv2}}
\newcommand{\cref}[2]{\hyperref[#2]{#1~\ref*{#2}}}
\newcommand{\colref}[2]{\hyperref[#2]{#1~\ref*{#2}}}

\newcommand{\figref}[1]{\colref{Figure}{#1}}

\newcommand{\secref}[1]{\colref{Section}{#1}}
\newcommand{\tabref}[1]{\colref{Table}{#1}}
\newcommand{\coloredref}[2]{\hyperref[#2]{#1~\ref*{#2}}}
\newcommand{\coloredsubref}[3]{\hyperref[#2]{#1~\ref*{#2}{#3}}}
\newcommand{\comment}[1]{}

\newcommand{\added}[1]{\textcolor{black}{#1}}

\defaultbibliographystyle{unsrtnat}
\defaultbibliography{refs}

\begin{document}
\begin{bibunit}

\begin{center}
\openup 0.5em
{
\usefont{OT1}{phv}{b}{n}\selectfont\Large{\added{\GRATEVtwo{}: Computational Tools for Real-time Analysis of High-Throughput High-Resolution TEM (HRTEM) Images of Conjugated Polymers}}
}

{\usefont{OT1}{phv}{m}{n}\selectfont\small
{Dhruv Gamdha$^1$, Ryan Fair$^3$, Adarsh Krishnamurthy$^{1,2}$, Enrique Gomez$^{3,4}$, Baskar Ganapathysubramanian$^{1,2}$*}}

{\usefont{OT1}{phv}{m}{n}\selectfont\small
{$^1$ Department of Mechanical Engineering, Iowa State University, Ames, IA\\
$^2$ Translational AI Research Center (TrAC), Iowa State University, Ames, IA\\
$^3$ Department of Chemical Engineering, The Pennsylvania State University, University Park, PA\\
$^4$ Department of Material Science and Engineering, The Pennsylvania State University, University Park, PA\\
* Corresponding Author Email: baskarg@iastate.edu
}}
\end{center}

\section*{Abstract}
Automated analysis of high-resolution transmission electron microscopy (HRTEM) images is increasingly essential for advancing research in organic electronics, where precise characterization of nanoscale crystalline regions is crucial for optimizing material properties. This paper introduces an open-source computational framework -- called \GRATEVtwo{}  (GRaph based Analysis of TEM, Version 2) -- designed for real-time analysis of HRTEM data, with a focus on characterizing complex microstructures in conjugated polymers, and illustrated using Poly[N-9’-heptadecanyl-2,7-carbazole-alt-5,5-(4’,7’-di-2-thienyl-2’,1’,3’-benzothiadiazole)] (PCDTBT), a key material in organic photovoltaics. \GRATEVtwo{} employs fast, automated image processing algorithms, enabling rapid extraction of structural features like d-spacing, orientation, and crystal shape metrics. Bayesian optimization rapidly identifies the parameters (that are traditionally user-defined) in the approach, reducing the need for manual parameter tuning and thus enhancing reproducibility and usability. Additionally, \GRATEVtwo{} is compatible with high-performance computing (HPC) environments, allowing for efficient, large-scale data processing at near real-time speeds. A unique feature of \GRATEVtwo{} is a Wasserstein distance-based stopping criterion, which optimizes data collection by determining when further sampling no longer adds statistically significant information. This capability optimizes the amount of time the TEM facility is used while ensuring data adequacy for in-depth analysis. Open-source and tested on a substantial PCDTBT dataset, this tool offers a powerful, robust, and accessible solution for high-throughput material characterization in organic electronics.

\subsection*{Keywords}
HRTEM $|$ Real-time analytics $|$ Graph algorithms $|$ Stopping criteria $|$ Automated fitting of material specific parameters 


\section{Introduction}
\label{sec:introduction}

Microscopy has long been a cornerstone in materials science, offering a unique window into the microstructure and enabling scientists to study properties at various scales, from the millimeter down to the atomic level~\citep{Williams2009TEM,Cullis2003Microscopy}. By visualizing otherwise inaccessible structures, microscopy provides critical insights into how atomic and molecular arrangements influence key macroscopic properties such as mechanical strength, electrical conductivity, and chemical reactivity~\citep{Egerton2009ElectronMicroscopy}. This fundamental understanding is essential for designing advanced materials with tailored properties for applications in fields such as electronics, energy storage, and catalysis.

High-resolution transmission electron microscopy (HRTEM) has advanced significantly in recent years, transforming nanoscale imaging and allowing researchers to capture atomic-level details of materials~\citep{Spence2013HighResolutionTEM}. HRTEM can now achieve sub-angstrom spatial resolution, making it possible to directly observe atomic lattices, defects, and interfaces that govern a material's behavior~\citep{Erni2009AtomicResolution}. The development of automated data acquisition systems has further broadened HRTEM's capabilities, enabling the collection of extensive datasets comprising hundreds or even thousands of high-resolution images~\citep{Kirkland2010AdvancedTEM}. These technological advances have opened new avenues for studying complex materials, such as organic semiconductors and conjugated polymers, with applications in organic electronics and photovoltaics~\citep{brabec2020material}.

While modern HRTEM has enhanced our ability to study complex materials, it also introduces challenges in data management and analysis due to the sheer volume and complexity of the data generated~\citep{Midgley2009ElectronTomography}. The scale of high-resolution datasets can quickly overwhelm traditional analysis workflows, which are often manual and time-consuming. This manual approach is highly dependent on the expertise and subjective judgment of the experimentalist, making it challenging to ensure consistency and reproducibility. In high-throughput applications where rapid feedback is essential, such as optimizing synthesis conditions or tracking real-time structural changes, the limitations of manual analysis become particularly pronounced.

To address these challenges, automated methods for HRTEM data analysis have emerged over the past decade. These methods aim to extract quantitative structural information from digital micrographs with minimal human intervention, improving both the efficiency and reliability of analysis~\citep{Egerton2009ElectronMicroscopy,TOTH2013TEMImageAnalysis, Zhu2017TEMImageAnalysis, Sharma1999FringeAnalysis, toth2021nanostructure}. Automated approaches can be broadly categorized into offline (post-acquisition) and online (real-time) methods. Offline methods involve processing data after acquisition, which is useful for extracting detailed structural information but may be computationally intensive and unable to keep pace with data acquisition rates~\citep{TOTH2013TEMImageAnalysis, Sharma1999FringeAnalysis, toth2021nanostructure,balaji-2019-CMS}. A motivating example is our prior work -- \texttt{GRATE} (GRaph based Analysis of TEM images) \cite{balaji-2019-CMS} -- that is based on Matlab. The computational and memory overhead makes offline methods less suitable for high-throughput applications where rapid processing is needed. The reliance on post-acquisition processing has underscored the need for integrating HRTEM with high-performance computing (HPC) resources to handle large volumes of data efficiently.

\begin{figure}[ht!]
    \centering
    \begin{subfigure}{\textwidth}
        \centering
        \begin{adjustbox}{max width=0.8\textwidth}
        \begin{tikzpicture}[
            node distance=0.5cm and 1cm,
            >=LaTeX,
            every node/.style={font=\large},
            process/.style={
                rectangle, draw=black, rounded corners, fill=blue!20,
                align=center, minimum width=3cm, minimum height=1cm
            },
            io/.style={
                trapezium, trapezium left angle=70, trapezium right angle=110,
                draw=black, fill=green!20, align=center,
                minimum width=2cm, minimum height=1cm
            },
            decision/.style={
                diamond, aspect=1.5, draw=black, fill=yellow!20,
                align=center
            },
            startstop/.style={
                rectangle, rounded corners, draw=black, fill=gray!20,
                align=center, minimum width=3cm, minimum height=1cm
            },
            arrow/.style={-{Latex[length=2mm]}, thick},
            ]
    
            \node (start) [startstop] {Start};

            \node (input) [io, below=of start] {
                \includegraphics[width=3.0cm]{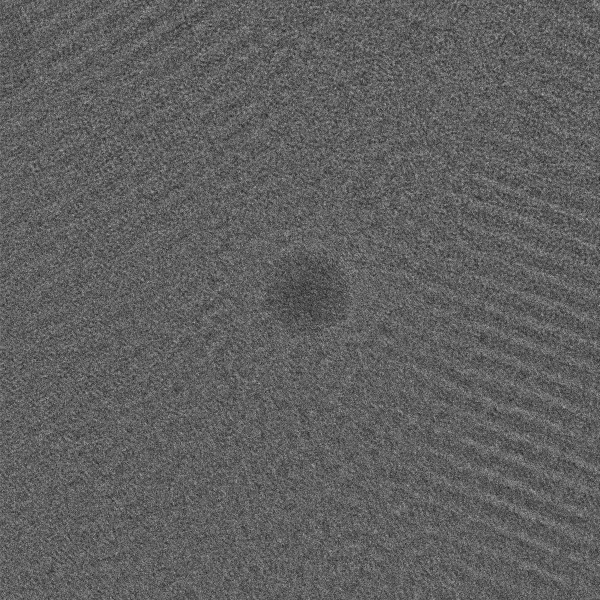} \\[1ex]
                Raw PCDTBT HRTEM Images 
            };

            \node (preprocessing) [process, below=of input] {{\Large \faImages[regular]} \\[1ex] Preprocessing};
            \node (processing) [process, below=of preprocessing] {{\Large \faIcon{cogs}} \\[1ex] Automated Image\\ Processing Algorithm};
            \node (output) [io, below=of processing] {
                \includegraphics[width=4.0cm]{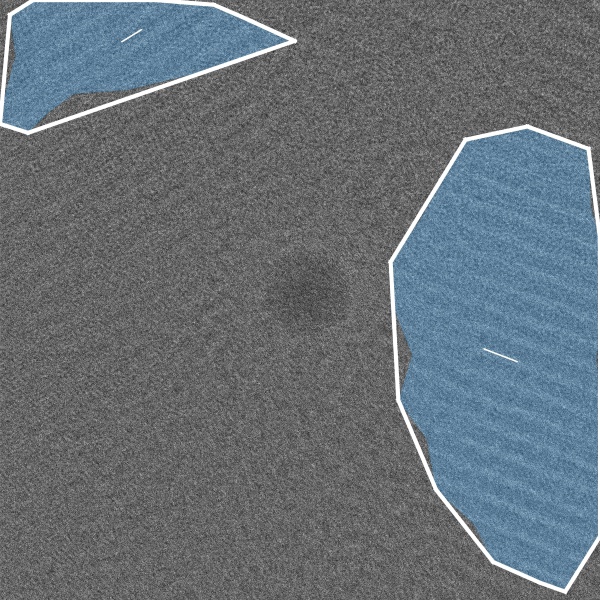} \\[1ex]
                {\Large \faChartBar\ \faChartArea\ \faChartPie} \\[1ex] Extracted Structural Features
                };
            
            \node (bayesopt) [process, right=3.0cm of processing] {{\Large \faIcon{brain}} \\[1ex] Bayesian Optimization\\ (Hyperparameters Tuning)};

            \node (gt_annotations) [io, above=of bayesopt] {
                \includegraphics[width=4.0cm]{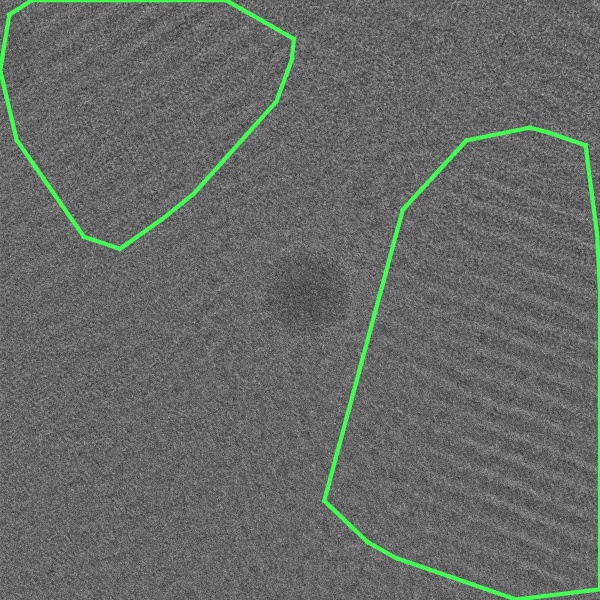} \\[1ex]
                Ground Truth Annotations
                };
    
            \node (datasuff) [process, left=3.0cm of processing] {{\Large \faCalculator\ \faDatabase } \\[1ex] Data Sufficiency\\ Criterion};

            \node (decision) [decision, below=of datasuff] {{\Large \faIcon{question-circle}} \\[1ex] Data\\ Sufficient?};

            \node (end) [startstop, below=of decision] {End};
    
            \draw [arrow] (start) -- (input);
            \draw [arrow] (input) -- (preprocessing);
            \draw [arrow] (preprocessing) -- (processing);
            \draw [arrow] (processing) -- (output);
    
            \draw [arrow, dashed] (bayesopt.west) -- node[above]{\scriptsize Optimized Hyperparameters} (processing.east);
            \draw [arrow, dashed] (processing.east) -- node[below]{\scriptsize Hyperparameters Updates} (bayesopt.west);
            
            \draw [arrow] (gt_annotations) -- (bayesopt);

            \draw [arrow] (processing) -- (datasuff);
    
            \draw [arrow] (datasuff) -- (decision);
    
            \draw [arrow] (decision.south) -- node[left]{Yes} (end.north);
            \draw [arrow] (decision.west) -- ++(-0.8cm,0) |- node[pos=0.25, left]{No} (input.west);
    
        \end{tikzpicture}
        \end{adjustbox}
        \caption{Schematic overview of the \GRATEVtwo{} computational framework.}
        \label{fig:framework_main}
    \end{subfigure}

    \vspace{1cm} 
    
    \begin{subfigure}{\textwidth}
        \centering
        \begin{adjustbox}{max width=0.6\textwidth}
        \begin{tikzpicture}[
            node distance=0.7cm and 1cm,
            >=LaTeX,
            every node/.style={font=\small},
            processPartA/.style={
                rectangle, draw=black, rounded corners, fill=blue!20,
                align=center, minimum width=2.5cm, minimum height=1cm
            },
            process/.style={
                rectangle, rounded corners, draw=black, fill=orange!20,
                align=center, minimum width=2.5cm, minimum height=0.8cm
            },
            io/.style={
                trapezium, trapezium left angle=70, trapezium right angle=110,
                draw=black, fill=green!20, align=center,
                minimum width=2.5cm, minimum height=0.8cm
            },
            arrow/.style={-{Latex[length=2mm]}, thick},
            ]
            \pgfdeclarelayer{background}
            \pgfsetlayers{background,main}
            
            \node (evaluation) [process] at (0,0) {Compute Objective\\ (e.g., $-\text{IoU}$)};
            \node (gt_annotations) [io, above=of evaluation] {
                \includegraphics[width=1.0cm]{figs/train_GT_FoilHole_21830243_Data_21829764_21829765_20200122_1102.jpg} \\[1ex]
                Ground Truth\\ Annotations
                };
            
            \node (image_processing) [processPartA, left=of evaluation] {{\large \faIcon{cogs}} \\[1ex] Automated Image\\ Processing Algorithm};
            
            \node (trainingData) [io, above=of image_processing] {
                \includegraphics[width=1.0cm]{figs/train_input_FoilHole_21830243_Data_21829764_21829765_20200122_1102.jpg} \\[1ex]
                Training Input
                };
            
            \node (surrogate_model) [process, below=of evaluation] {Update Surrogate\\ Model (GP)};
            \node (acquisition) [process, below=of surrogate_model] {Optimize Acquisition\\ Function};
            \node (next_params) [process, below=of acquisition] {Select Next\\ Hyperparameters};
            
            \draw [arrow] (gt_annotations) -- (evaluation);
            \draw [arrow] (image_processing) -- (evaluation);
            \draw [arrow] (evaluation) -- (surrogate_model);
            \draw [arrow] (surrogate_model) -- (acquisition);
            \draw [arrow] (acquisition) -- (next_params);
            \draw [arrow] (trainingData) -- (image_processing);
            
            \coordinate (loop_start) at ($(next_params.south) + (0,-0.3cm)$);
            \coordinate (loop_left) at ($(loop_start) + (-3.90cm,0)$);
            \coordinate (loop_up) at ($(image_processing.south) + (0,-0.5cm)$);

            \draw [arrow, dashed] (next_params.south) -- (loop_start) -- (loop_left) -- node[left, align=center]{{\scriptsize Hyperparameters} \\ {\scriptsize Updates Loop}}(loop_up) -- (image_processing.south);
            
            \node (optimized_params) [process, right=of next_params, xshift=1cm] {
            {\large \faList*[regular]\ \faCheck} \\[1ex] Optimized\\ Hyperparameters
            };
            \draw [arrow] (next_params) -- (optimized_params);
            
            \coordinate (bo_loop_nw) at ($(evaluation.north west) + (-0.5cm, 0.3cm)$);
            \coordinate (bo_loop_se) at ($(next_params.south east) + (0.5cm, -0.5cm)$);
            
            \begin{pgfonlayer}{background}
                \node (bo_loop) [draw, dashed, fit=(bo_loop_nw)(bo_loop_se), fill=blue!10] {};
            \end{pgfonlayer}
            
            \node[right=2.0cm of bo_loop.north,
                  anchor=north west,
                  align=left] {
                    {\large Bayesian Optimization} \\ (\normalsize Hyperparameter Tuning)
                    };
            
        \end{tikzpicture}
        \end{adjustbox}
        \caption{Detailed diagram of the Bayesian Optimization process.}
        \label{fig:bayes_opt}
    \end{subfigure}

    \caption{Schematic overview of the \GRATEVtwo{} computational framework. The framework processes raw HRTEM images of PCDTBT, applies preprocessing, and performs automated image processing with parameters optimized via Bayesian optimization. A data sufficiency criterion based on the Wasserstein distance assesses whether additional TEM data is needed. The output comprises extracted structural features such as d-spacing, orientation, and crystal shape metrics. (a) The overall computational framework of \GRATEVtwo{}, and (b) the detailed Bayesian Optimization (BO) component used for parameter tuning.}
    \label{fig:overall_framework_two_parts}
\end{figure}

\begin{figure}[t!]
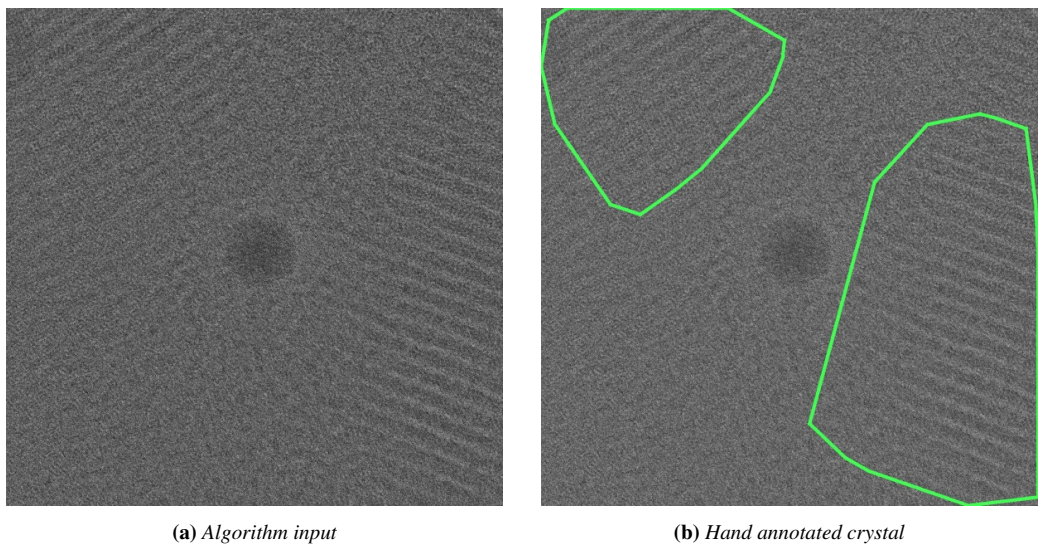

    \centering
    \begin{subfigure}[b]{0.4\linewidth}
        \centering
        \includegraphics[width=\linewidth]{figs/train_input_FoilHole_21830243_Data_21829764_21829765_20200122_1102.jpg}
        \caption{Algorithm input}
        \label{fig:OriginalImg}
    \end{subfigure}
    \hspace{0.02\linewidth}
    \begin{subfigure}[b]{0.4\linewidth}
        \centering
        \includegraphics[width=\linewidth]{figs/train_GT_FoilHole_21830243_Data_21829764_21829765_20200122_1102.jpg}
        \caption{Hand annotated crystal}
        \label{fig:annotatedImg}
    \end{subfigure}
    \caption{Comparison between (a) the original HRTEM image and (b) the manually annotated crystals used for training the Bayesian model. 13 manually annotated images are used for effective training in the work. Manual annotation is time-consuming and subjective.}
    \label{fig:ManualAnnotationCrystal}
\end{figure}

In recent years, there has been a strong push toward developing \textit{in situ}, or real-time, automated analysis methods~\citep{pratiush2024realizing, ghosh2024integrating, pratiush2024realizingSTEM}. \textit{In situ} analysis allows for data interpretation during experiments, providing immediate feedback that can guide adjustments in experimental parameters. This capability is particularly valuable in dynamic experiments, such as observing structural changes in response to external stimuli or monitoring materials during synthesis~\citep{tsarfati2024charge, tsarfati2023structural, bi2023situ, winkler2024texture, da2022assessment, chao2023situ, tsarfati2022microstructural}. Real-time analysis requires methods that can handle high-resolution data quickly and accurately, maintaining performance under the demanding conditions of live data acquisition.

Machine learning (ML) has shown promise in automating feature recognition in HRTEM images, with deep learning models capable of identifying patterns and structures in high-resolution data~\citep{masud2024machine, Ziatdinov2017DeepLearningTEM, catherine-2021-MicroscopyAndMicroanalysis}. However, ML approaches have limitations in this context. They typically require large, annotated datasets for training, which is resource-intensive and time-consuming to create~\citep{rade20223d, Ziatdinov2017DeepLearningTEM, Azimi2018MachineLearningMaterials}. Each chemical system may exhibit unique structural features and imaging artifacts, requiring retraining or fine-tuning of models to ensure accuracy~\citep{TOTH2013TEMImageAnalysis, toth2021nanostructure}. This dependence on extensive training data reduces the adaptability of ML methods across different material systems, particularly in high-throughput or exploratory research where materials and imaging conditions vary frequently. Recent work in creating foundational models, and machine learning approaches like transfer learning and few-shot learning applied to these problems show promise.

In contrast, robust image processing techniques provide a more flexible and efficient solution for automated analysis. Image processing-based approaches can generate interpretable, reproducible data in real time without the need for extensive, system-specific training. These methods typically involve well-established procedures like frequency filtering, binarization, and morphological operations that are adaptable across diverse materials~\citep{Sharma1999FringeAnalysis, Rouzaud2002ImageAnalysis, TOTH2013TEMImageAnalysis, Palotas1996Microscopy, Yehliu2011CombustionFlame, balaji-2019-CMS}. In conjunction with ML models, image processing methods offer a practical solution for rapid, in situ data analysis with minimal human input, making them particularly valuable for high-throughput studies in materials science.

\GRATEVtwo{} aims to bridge the gap between offline and real-time analysis in high-resolution transmission electron microscopy (HRTEM) by providing an automated, image-processing-based framework with minimal human intervention. Tailored for high-throughput settings, \GRATEVtwo{} combines rapid data extraction with a robust and user-friendly parameter optimization process. By focusing on image processing techniques augmented with Gaussian process optimization, \GRATEVtwo{} minimizes the need for manual parameter selection and tuning, enhancing reproducibility and accessibility for researchers.

A schematic overview of the \GRATEVtwo{} computational framework is shown in \figref{fig:framework_main}. The framework processes raw HRTEM images of PCDTBT (\figref{fig:OriginalImg}), applies preprocessing, and performs automated image processing with parameters optimized via Bayesian optimization (\figref{fig:bayes_opt}). Traditionally, manual annotation of crystals in HRTEM images is time-consuming and subjective (\figref{fig:ManualAnnotationCrystal}). In our approach, only a dozen manually annotated images (\figref{fig:annotatedImg}) are used as input to a  Bayesian optimizer to rapidly identify material-specific image processing parameters. This minimal requirement significantly reduces the burden on researchers, facilitating rapid deployment of the algorithm on new datasets.

\pagebreak

\GRATEVtwo{} introduces several key innovations to address the limitations of current HRTEM analysis methods:
\begin{enumerate}
    \item \GRATEVtwo{} offers fast and efficient processing of HRTEM images on the order of a few seconds per image. By supporting batch processing and multiprocessing, it enables rapid analysis of large datasets, significantly improving throughput.
    \item It extends advanced image analysis techniques to organic polymer materials, specifically PCDTBT, which have not been extensively studied with these methods.
    \item Bayesian optimization is employed within \GRATEVtwo{} to automate the tuning of material-specific image processing parameters, allowing the framework to adapt to different datasets with minimal expert input (\figref{fig:bayes_opt}).
    \item The algorithm parameters are constructed as functions of known d-spacing values, simplifying parameter selection and making the method more accessible to users without extensive image processing expertise. This also ensures that parameter selection is interpretable and, thus, scientifically justified.
    \item \GRATEVtwo{} incorporates a data sufficiency criterion based on the Wasserstein distance to guide data collection efforts. This criterion provides a quantitative stopping point, indicating when further TEM data collection no longer yields additional insights, which is important when access to imaging facilities is limited and imaging is expensive.
\end{enumerate}

By optimizing data collection, \GRATEVtwo{} helps experimentalists avoid unnecessary resource expenditure while ensuring data quality. This combination of fast, automated processing, real-time adaptability, and efficient data collection positions \GRATEVtwo{} as a powerful tool for materials research, particularly in the study of organic electronic materials such as conjugated polymers~\citep{Park2020HighEfficiencyOPV, Rodriguez2018PCDTBTAmorphous, Beiley2011PCDTBTSemicrystalline}.

\section{Results}
\label{sec:results}
We first provide brief details of the material and process used to collect the HRTEM data for completeness. To prepare TEM samples, $5~mg/mL$ solutions of PCDTBT were dissolved in chlorobenzene within a nitrogen glovebox at $45^{\circ}\mathrm{C}$ for at least 12 hours. Silicon wafers were cleaned by sonication for 20 minutes in acetone, followed by 20 minutes in isopropanol, and subsequently subjected to UV-ozonation for 20 minutes. Poly(3,4-ethylenedioxythiophene)-poly(styrene sulfonate) (PEDOT:PSS) films were cast onto silicon substrates by spin coating at 4000 RPM for 2 minutes in air to serve as a sacrificial layer, facilitating film floating. Substrates were then transferred to a nitrogen glovebox, where PCDTBT films were spin-cast at 800 RPM for 2 minutes. For TEM sample preparation, the coated substrates were removed from the glovebox, floated in deionized water, and carefully transferred onto copper TEM grids. These samples were left under ambient conditions to dry overnight and then annealed in the nitrogen glovebox at 190 °C for 2 hours. High-resolution TEM (HRTEM) imaging was performed at the Penn State Materials Characterization Lab using the FEI Titan Krios microscope operating at 300 kV, equipped with the K2 direct electron detector and a cryo-stage. A dose rate of 20 e/Å²s was used for a 2.5 s exposure. An automated acquisition process was set to autofocus before each capture at 300 kx magnification, with a randomly assigned defocus value between 0 and -3 µm. Images were acquired at 470 kx magnification with a 2.5 µm step size between exposed regions, resulting in a total of 637 images for subsequent analysis. 

Using \GRATEVtwo{}, we detected crystals within each of the HRTEM images. For each identified crystal, the following features were extracted (see \secref{sec:results_detection}): center-of-mass coordinates, orientation angle relative to the image axis, d-spacing, and crystal lengths along both the major and minor axes. Through this process, \GRATEVtwo{} identified a total of 4350 ordered domains from the HRTEM images. In \secref{sec:results_timing}, we present the timing performance of \GRATEVtwo{}, along with the time distribution among different components of the analysis process. In \secref{sec:results_dataSufficiency}, we evaluate the data sufficiency—that is, how many images are sufficient to achieve statistical convergence in the extracted features.

\subsection{Advantages of Bayesian Parameter Optimization}
\label{sec:results_bayesian_optimization}

To evaluate the effectiveness of integrating Bayesian optimization for hyperparameter tuning in our image processing algorithm, we conducted a comparative analysis between manually selected parameters and the Bayesian-optimized parameters. The performance was assessed using the Intersection over Union (IoU) metric between the algorithm's output against expert-annotated ground truth on six representative HRTEM images.

\paragraph{Quantitative Comparison of IoU Scores}

\tabref{table:iou_scores} presents the IoU scores for each image using both the manually selected parameters and the Bayesian-optimized parameters. Additionally, the table includes the differences in IoU scores between the two parameter sets for each image. The average and standard deviation of IoU score across the six images is also calculated for both cases.

\begin{table}[t!]
    \caption{IoU Scores over validation dataset for Manual and Bayesian-Optimized Parameters with Differences}
    \label{table:iou_scores}
    \setlength{\extrarowheight}{2pt}
    \centering
    \begin{tabular}{||c|c|c|c||}
    \hline
    \textbf{Image Filename} & \textbf{Manual Selected} & \textbf{Bayesian} & \textbf{IoU Difference, $d_i$} \\
    & \textbf{Parameters IoU} & \textbf{Parameters IoU} &  \\
    \hline\hline
    \texttt{1.tif} & 0.5296 & 0.6911 & +0.1615 \\
    \hline
    \texttt{2.tif} & 0.3274 & 0.5156 & +0.1882 \\
    \hline
    \texttt{3.tif} & 0.3178 & 0.4934 & +0.1756 \\
    \hline
    \texttt{4.tif} & 0.5285 & 0.5864 & +0.0579 \\
    \hline
    \texttt{5.tif} & 0.3906 & 0.5400 & +0.1494 \\
    \hline
    \texttt{6.tif} & 0.5288 & 0.6438 & +0.1150 \\
    \hline \hline
    \textbf{Average IoU} & \textbf{0.4371} & \textbf{0.5784} & \textbf{+0.1413} \\
    \hline
    \end{tabular}
\end{table}

The average IoU score using the Bayesian-optimized parameters is 0.5784, representing an improvement of approximately 32.3\% over the average IoU of 0.4371 obtained with the manually selected parameters. This significant increase demonstrates the effectiveness of Bayesian optimization in enhancing the algorithm's performance in detecting and segmenting crystalline regions in HRTEM images. The computed t-value corresponding to t-statistic is $7.074$, which exceeds the critical t-value of $2.571$, indicating strong statistical justification for improvement.

\paragraph{Visual Comparison of Detection Results}

\begin{figure}[t!]
    \centering
    \begin{tabular}{>{\centering\arraybackslash}m{0.3\textwidth} 
                    >{\centering\arraybackslash}m{0.3\textwidth} 
                    >{\centering\arraybackslash}m{0.3\textwidth}}
        \toprule
        \textbf{Ground truth (i.e. expert hand identifying crystals)} & \textbf{Results using manually selected parameters} & \textbf{Results using Bayesian-Optimized parameters} \\
        \midrule
        \includegraphics[width=\linewidth]{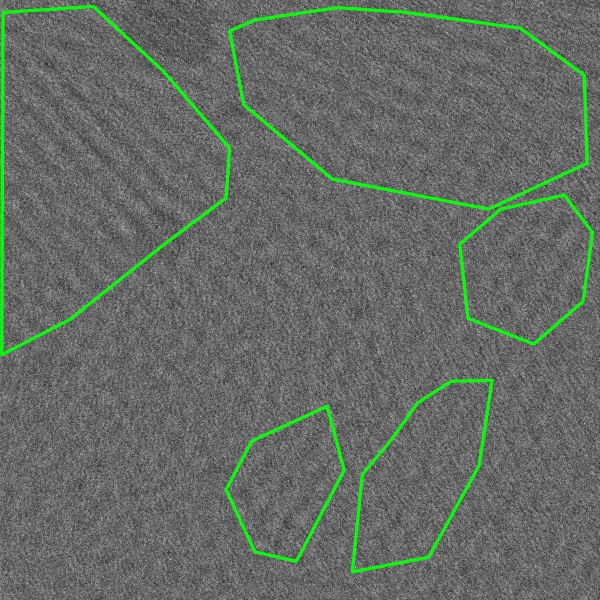} &
        \includegraphics[width=\linewidth]{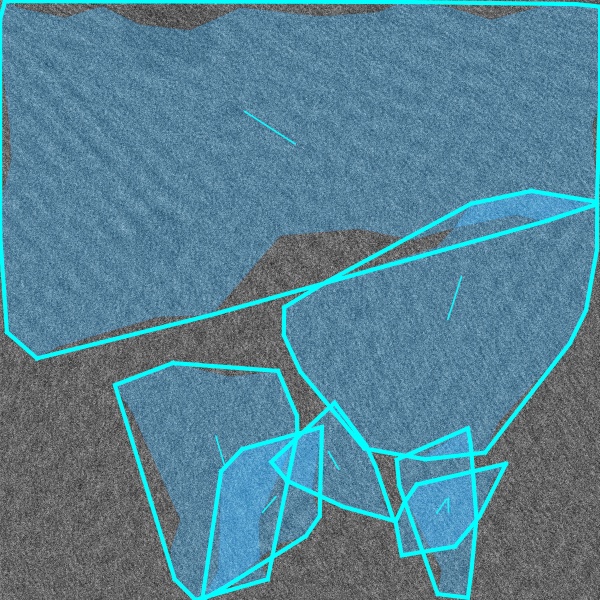} &
        \includegraphics[width=\linewidth]{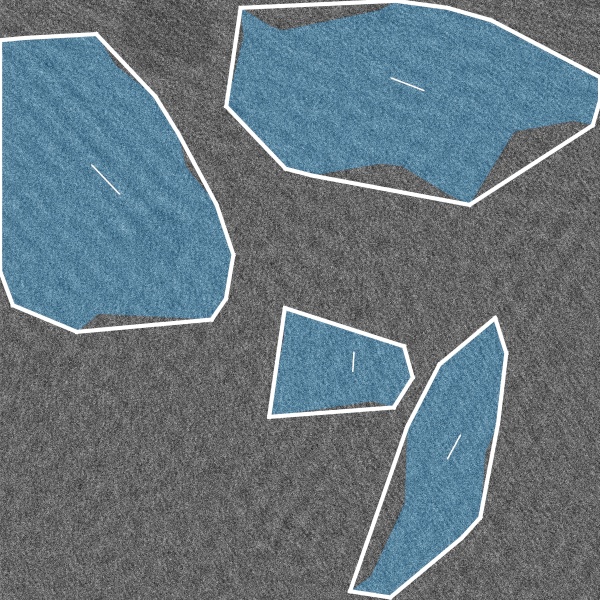} \\
        \midrule
        \includegraphics[width=\linewidth]{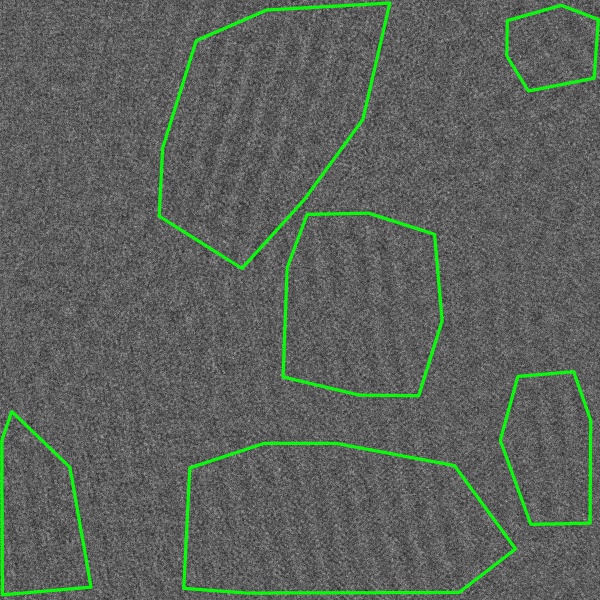} &
        \includegraphics[width=\linewidth]{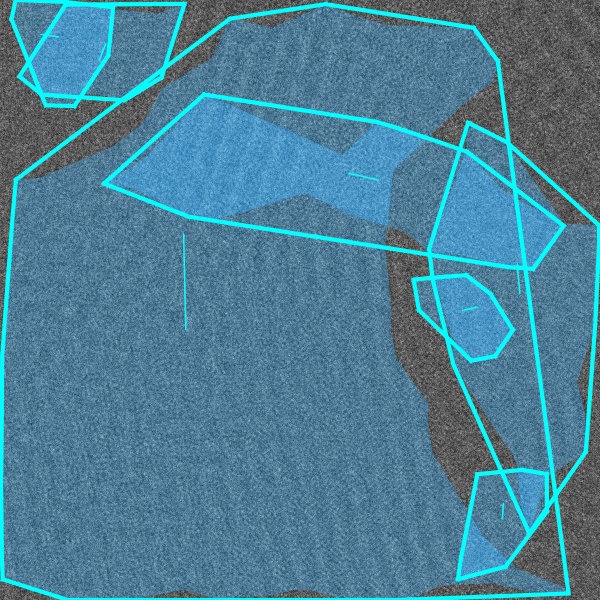} &
        \includegraphics[width=\linewidth]{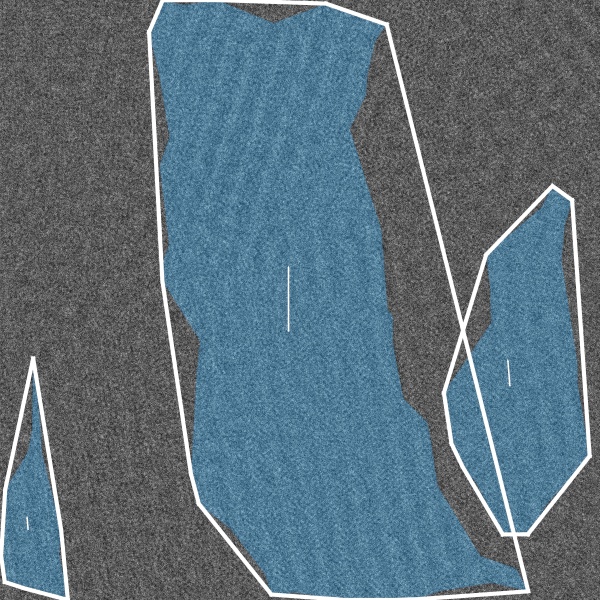} \\
        \midrule
        \includegraphics[width=\linewidth]{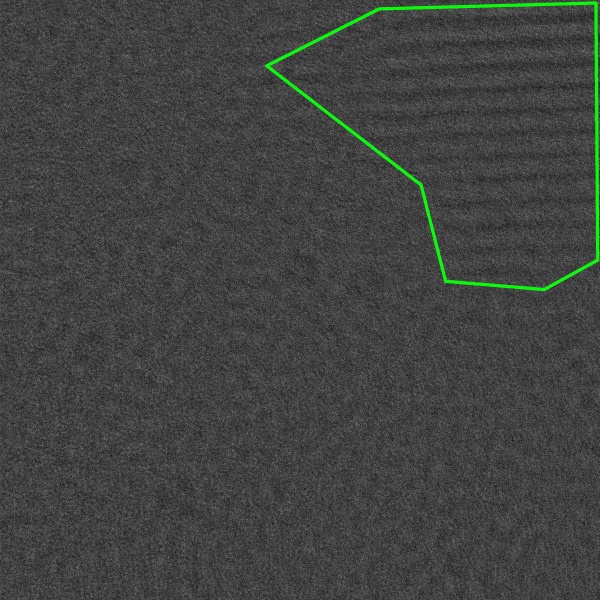} &
        \includegraphics[width=\linewidth]{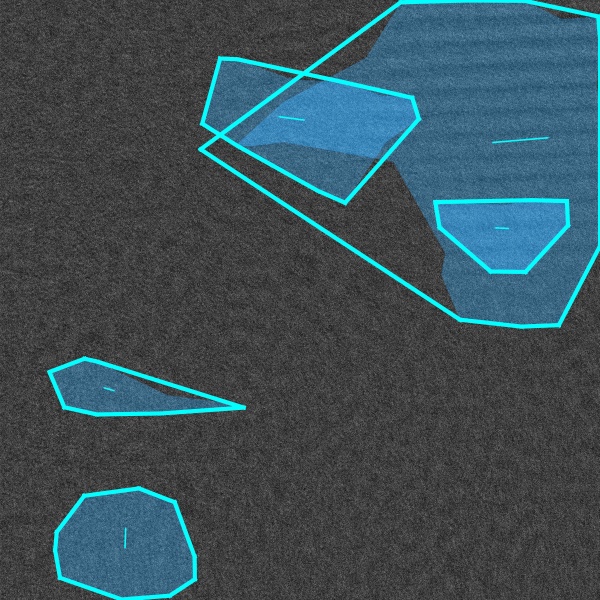} &
        \includegraphics[width=\linewidth]{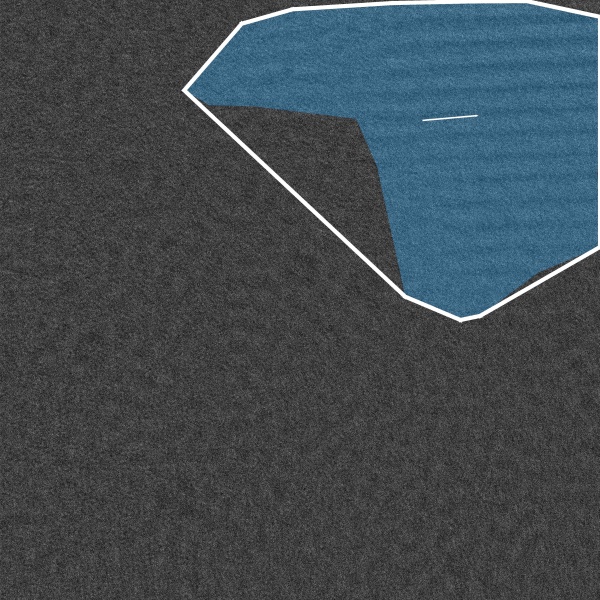} \\
        \bottomrule
    \end{tabular}
    \caption{Comparison of Ground Truth, Manually selected parameter detection, and Bayesian-Optimized parameter detection across three different images. Each column represents a distinct detection method, illustrating how Bayesian optimization enhances segmentation accuracy by more closely matching the ground truth annotations compared to manual parameter tuning. The Bayesian optimization process was conducted over \textbf{200 iterations}, achieving a minimum loss value of \textbf{-0.7319} at the \textbf{151st evaluation}.}
    \label{fig:comparison_manual_bayesian_grid}
\end{figure}

\figref{fig:comparison_manual_bayesian_grid} and \cref{Appendix}{appendix:AdditionalValidationResults} provide qualitative comparisons between the segmentation results obtained with manually selected parameters and those obtained with Bayesian-optimized parameters. Visual comparisons illustrate that the algorithm using Bayesian-optimized parameters generally provides a more accurate and comprehensive detection of crystalline regions than the manually tuned approach. The Bayesian-optimized parameters tend to align the detected crystal boundaries more consistently with the annotated features. In particular, the algorithm with Bayesian-optimized parameters appears to capture finer details and produce clearer delineation of crystal boundaries, suggesting a closer correspondence to the expert annotations than that achieved under manual parameter tuning.

\paragraph{Convergence of the Bayesian Optimization Process}

\begin{figure}[ht]
    \centering
    \begin{tikzpicture}
        \begin{axis}[
            width=0.6\linewidth,                         
            height=0.4\linewidth,                         
            xlabel={Iterations},               
            ylabel={Minimum Loss},              
            grid=major,                         
            grid style={dashed, gray!30},       
            legend style={anchor=north east},   
            ]
            \addplot[
                color=blue, 
                mark=*,     
                ] 
                table [
                    x=Evaluations,                  
                    y=Min Objective Value,          
                    col sep=comma,                  
                ] {data/BO_run4_convergence.csv};   
            \addlegendentry{Min Objective Value}    
        \end{axis}
    \end{tikzpicture}
    \caption{Convergence of the Bayesian optimization process over 200 Iterations, illustrating the reduction in the loss function (negative IoU) as optimization progresses. The y-axis represents the minimum loss value achieved up to that evaluation. The minimum loss value of -0.7319 was attained at the 151st evaluation.}
    \label{fig:loss_vs_evaluation}
\end{figure}
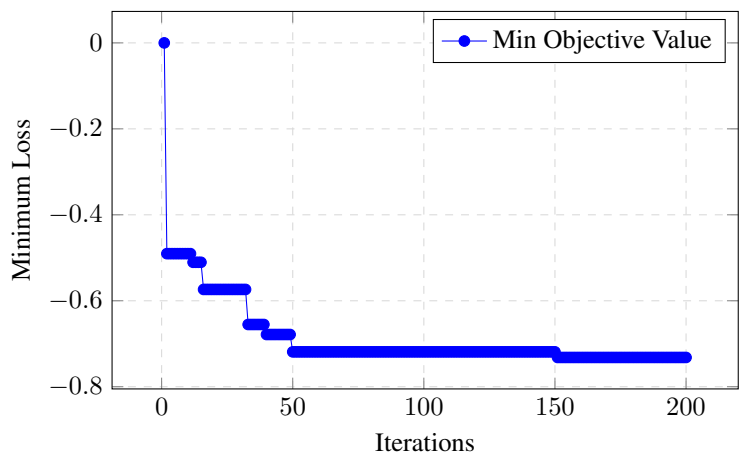

\figref{fig:loss_vs_evaluation} depicts the convergence of the Bayesian optimization process. The objective function, defined as the negative mean IoU, decreases over successive iterations, indicating that the optimization algorithm effectively identifies hyperparameters that enhance the segmentation performance.
The convergence plot demonstrates that significant improvements are achieved within the first 51 iterations, after which the objective function gradually approaches a plateau. 

\paragraph{Analysis of Optimized Parameters}

\cref{Appendix}{appendix:OptimumParameters} compares the values of the manually selected parameters with those obtained through Bayesian optimization.

Several notable differences are observed between the parameter sets:

\begin{itemize}
    \item \textbf{Morphological Operations}: The Bayesian-optimized parameters for \texttt{closing\allowbreak\_k\allowbreak\_size } and \texttt{opening\allowbreak\_k\allowbreak\_size } are significantly smaller than the manual values (2 vs. 15 and 17, respectively). This suggests that less aggressive morphological operations preserve finer details in the images, contributing to improved segmentation accuracy.
    \item \textbf{Edge Detection and Filtering}: Parameters like \texttt{ellipse\allowbreak\_len\allowbreak\_propCons} and \texttt{dspace\allowbreak\_bandpass} are adjusted to better capture the characteristics of the crystalline structures. The increase in \texttt{ellipse\allowbreak\_len\allowbreak\_propCons} from 1.5 to 4.03 indicates a preference for detecting longer ellipses, aligning with the elongated shapes of crystals.
    \item \textbf{Thresholding Parameters}: The \texttt{pixThresh\allowbreak\_propCons} is slightly higher in the Bayesian-optimized parameters, which may help in differentiating crystals from the background noise more effectively.
\end{itemize}

\paragraph{Advantages of Bayesian Optimization}

The integration of Bayesian optimization into our image processing framework offers several advantages:

\begin{enumerate}
    \item \textbf{Enhanced Performance}: The Bayesian-optimized parameters significantly improve the mean IoU score by approximately 32.3\%, indicating superior segmentation performance.
    \item \textbf{Automated Tuning}: Bayesian optimization automates the hyperparameter tuning process, reducing the need for manual intervention and deep image-processing expertise, thereby saving time and resources. This is especially important when the number of hyperparameters (here, 13) makes manual exploration rather tedious. 
    \item \textbf{Efficient Exploration}: The optimization process efficiently explores the hyperparameter space, converging to optimal values in fewer evaluations compared to exhaustive search methods. This is evident from the convergence plot in \figref{fig:loss_vs_evaluation}.
    \item \textbf{Robustness}: The optimized parameters produce robust results across various images, as evidenced by consistent improvements in IoU scores. 
\end{enumerate}

\subsection{Structural Feature Extraction of Crystals from HRTEM Images}
\label{sec:results_detection}

\begin{figure}[!b]
    \centering
    \begin{subfigure}[b]{0.49\linewidth}
        \centering
        \includegraphics[width=\linewidth]{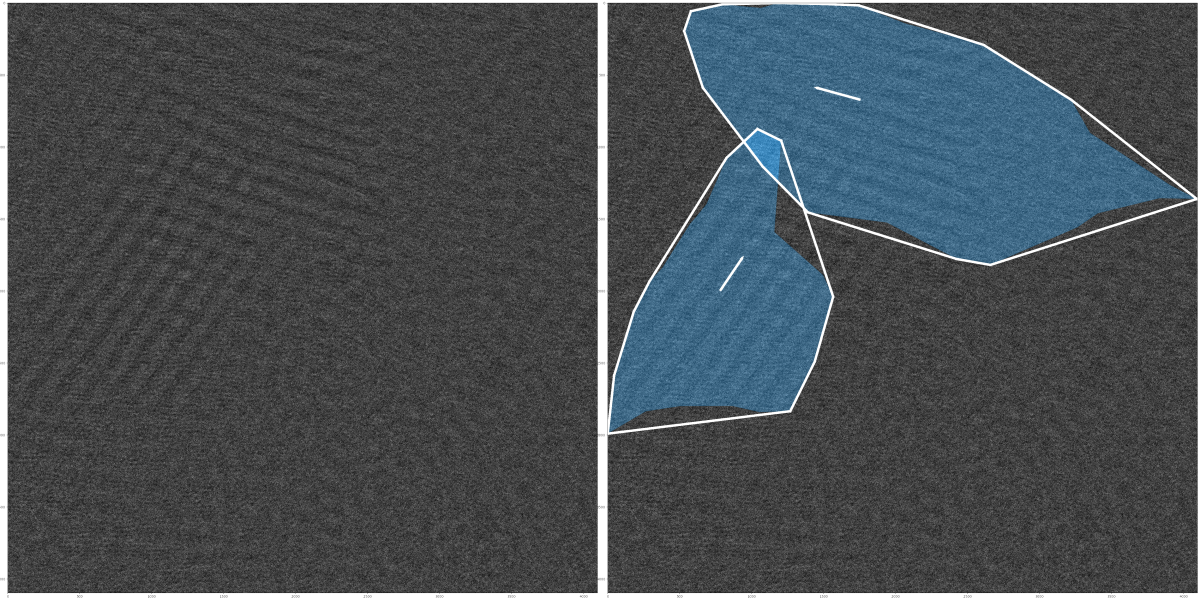}
        \caption{}
    \end{subfigure}
    \begin{subfigure}[b]{0.49\linewidth}
        \centering
        \includegraphics[width=\linewidth]{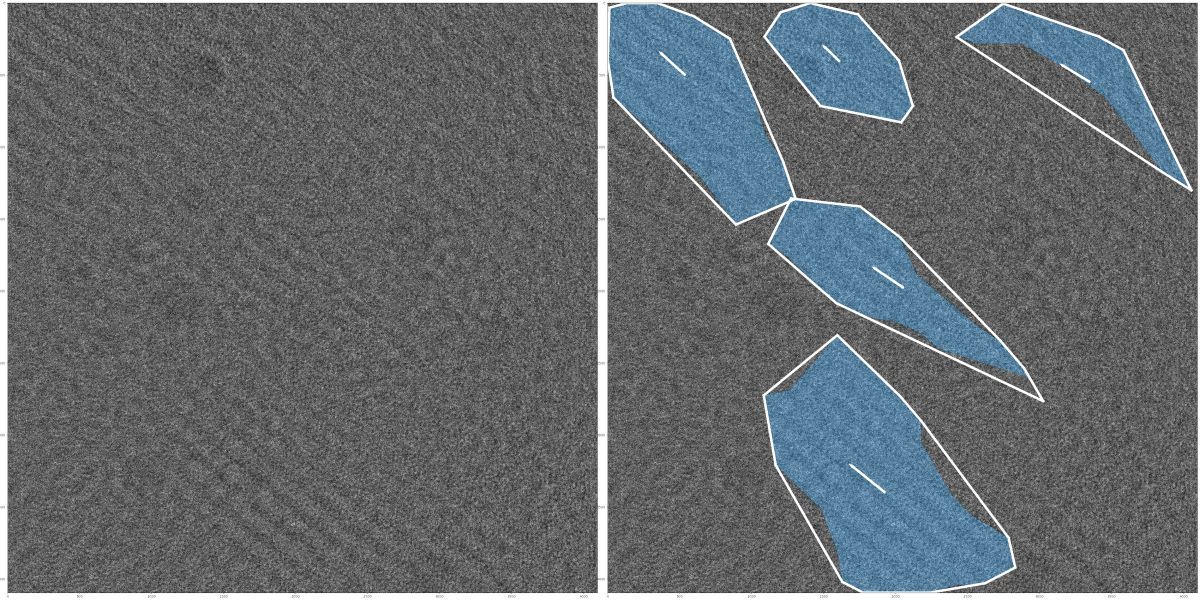}
        \caption{}
    \end{subfigure}
    \caption{(a) and (b) are 1.tif and 2.tif images respectively corresponding to \tabref{table:HRTEMDetectionResults} and \tabref{table:HRTEMCrystalCorrelation}. (a) and (b) shows the original image (left) and the segmentation output (right) from our algorithm for HRTEM of PCDTBT. The detected crystals have a d-spacing of 1.9nm. The image on the left is the input to the algorithm, and on the right is the output of the algorithm. Each detected crystal in the output shows (1) the convex hull boundary around the crystal, (2) the shaded region representing a more exact crystal region, and (3) a straight line at the centroid of the convex hull, which shows the orientation of the crystal patterns. }
    \label{fig:HRTEMDetectionResults}
\end{figure}

\begin{table}[t!]
    \caption{Features of the detected crystals shown in \figref{fig:HRTEMDetectionResults}.}
    \label{table:HRTEMDetectionResults}
    \small
    \setlength{\extrarowheight}{2pt}
    \centering
    \begin{tabular}{||c|c|c|c|c|c|c|c||} 
    \hline
    Name & Centroid & Area & Angle &  d-Spacing & MajorAxis & MinorAxis & AxisAngle \\
    & $(px, px)$ & $(nm^2)$ & $(deg)$ & $(nm)$ & $(nm)$ & $(nm)$ & $(deg)$ \\ [0.5ex]
    \hline\hline
    1.tif   & (1748 , 670)  & 589.7 & -164.7    & 2.1   & 21.1  & 10.2  & 23.8 \\
    \hline
            & (785  , 1992) & 293.9 & -55.9     & 2.0   & 14.8  & 9.9   & -65.5 \\
    \hline
    2.tif   & (534  , 497)  & 177.9 & -137.9    & 1.9   & 12.5  & 5.3   & 53.0 \\
    \hline
            & (1607 , 402)  & 84.3  & -136.0    & 0.8   & 7.9   & 4.3   & 35.4 \\
    \hline
            & (3345 , 546)  & 71.7  & -148.8    & 1.7   & 14.4  & 4.8   & 39.6 \\
    \hline
            & (2050 , 1975) & 125.4 & -146.0    & 2.2   & 16.0  & 4.2   & 34.1 \\
    \hline
            & (1922 , 3396) & 263.8 & -141.4    & 2.0   & 14.2  & 8.0   & 46.9 \\
    \hline
    \end{tabular}
\end{table}

\begin{table}[t!]
    \centering
    \small
    \caption{Crystal correlation measurements for \figref{fig:HRTEMDetectionResults}.}
    \label{table:HRTEMCrystalCorrelation}
    \setlength{\extrarowheight}{2pt}
    \begin{tabular}{||c|c|c|c||} 
    \hline
    Name & Metric Distance & Direct Distance & Relative Angle \\
    & $(1)$ & $(nm)$ & $(deg)$  \\ [0.5ex]
    \hline\hline
    1.tif   & 0.89 & 20.84 & 71.24 \\
    \hline
    2.tif   & 2.91 & 35.81 & 10.95 \\
    \hline
            & 1.95 & 26.97 & 8.13 \\
    \hline
            & 2.45 & 40.94 & 3.5 \\
    \hline
            & 2.21 & 24.57 & 2.81 \\
    \hline
            & 2.91 & 40.58 & 7.45 \\
    \hline
            & 1.17 & 18.18 & 4.64 \\
    \hline
    \end{tabular}
\end{table}

\begin{figure}[t!]
    \centering
    \begin{subfigure}[b]{0.32\linewidth} 
        \centering
        \includegraphics[width=\linewidth,trim={0.6in 0.6in 0.6in 0.6in}, clip]{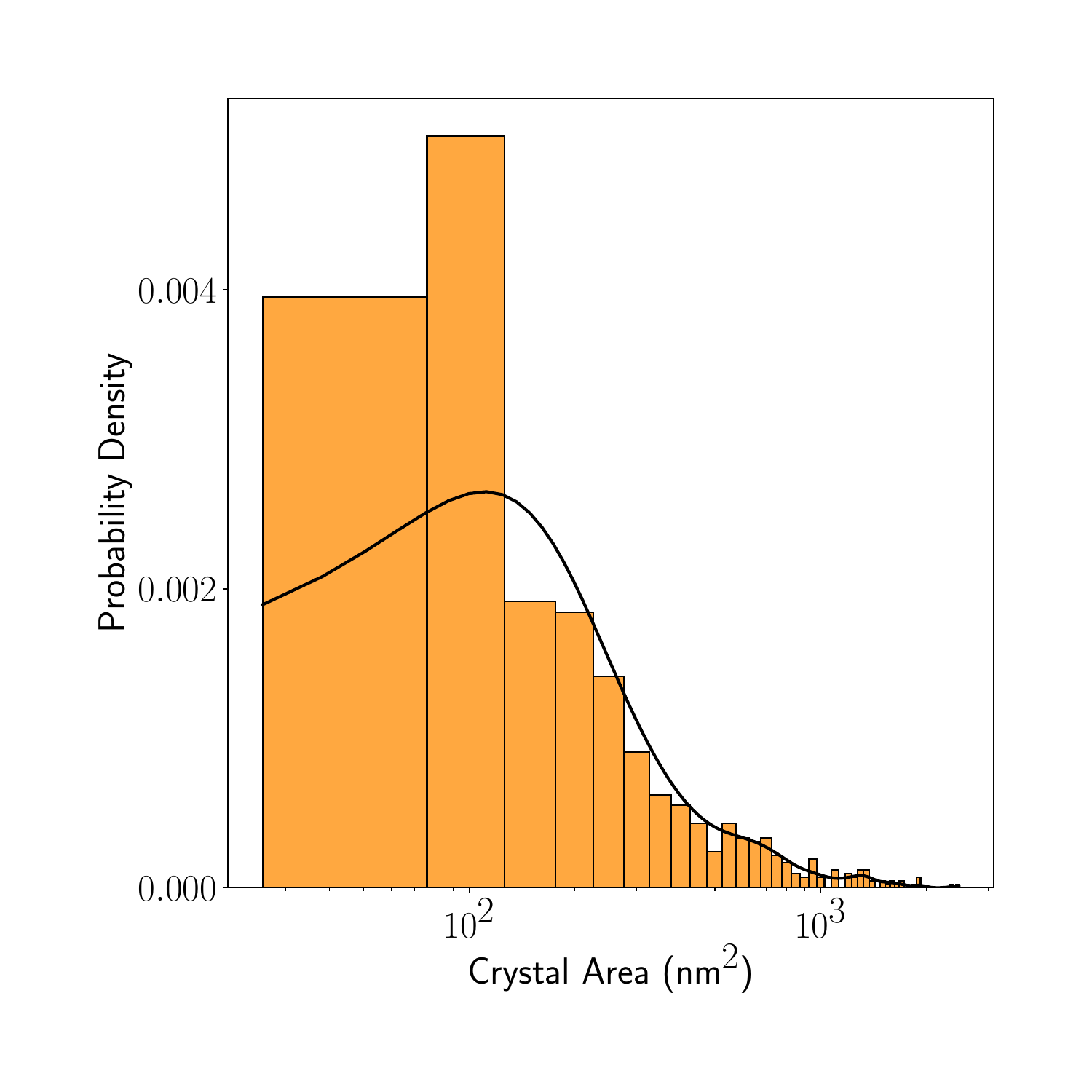}
        \caption{Crystal area}
        \label{fig:CrystalAreaHistogram}
    \end{subfigure} 
    \begin{subfigure}[b]{0.32\linewidth} 
        \centering
        \includegraphics[width=\linewidth,trim={0.6in 0.6in 0.6in 0.6in}, clip]{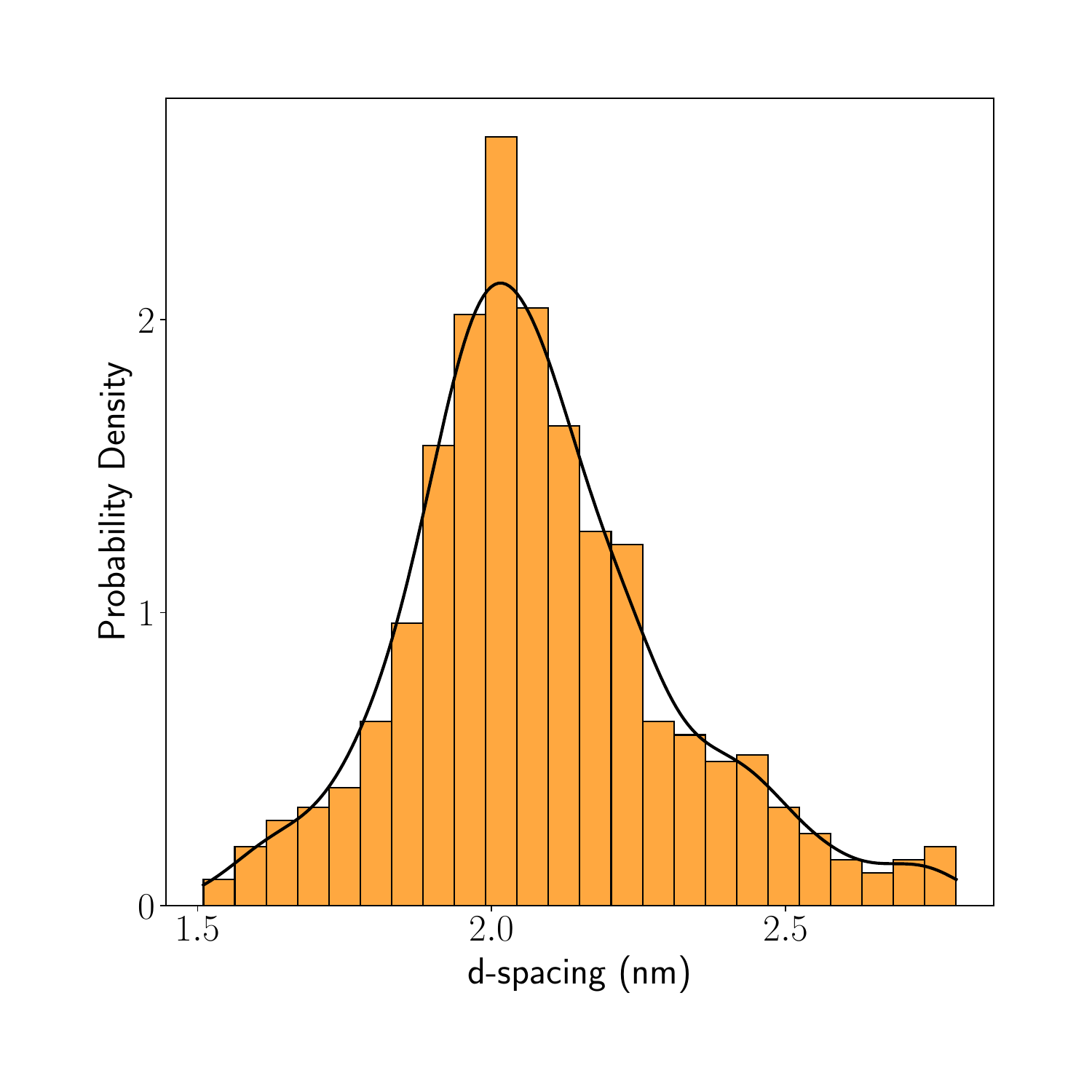}
        \caption{Crystal d-spacing}
        \label{fig:CrystalDSpacingHistogram}
    \end{subfigure}
    \begin{subfigure}[b]{0.32\linewidth} 
        \centering
        \includegraphics[width=\linewidth,trim={0.6in 0.6in 0.6in 0.6in}, clip]{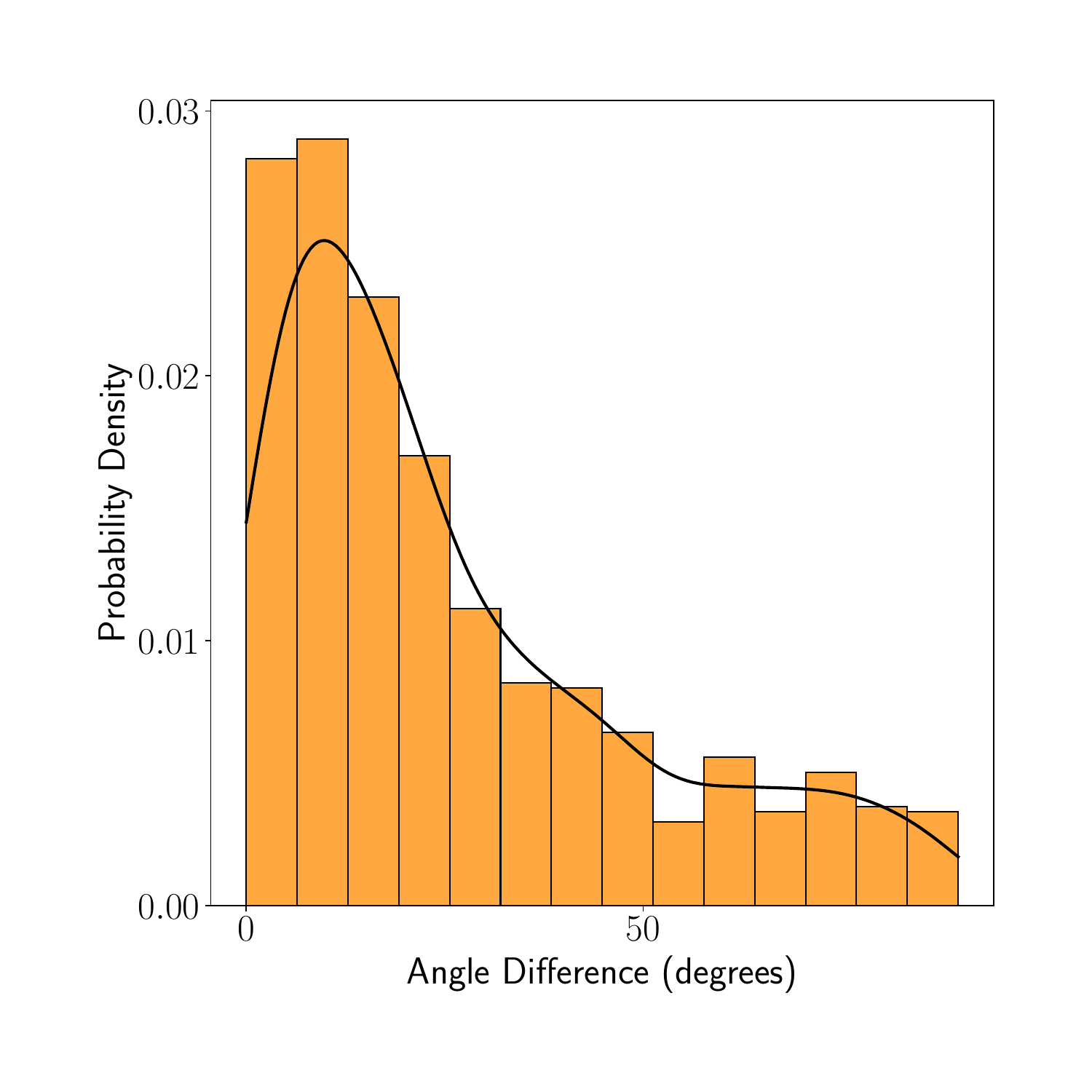}
        \caption{Angle difference}
        \label{fig:CrystalOrientationVsMajorAxisHistogram}
    \end{subfigure}\\
    \begin{subfigure}[b]{0.40\linewidth} \centering
        \centering
        \includegraphics[width=\linewidth,trim={0.6in 0.6in 0.6in 0.6in}, clip]{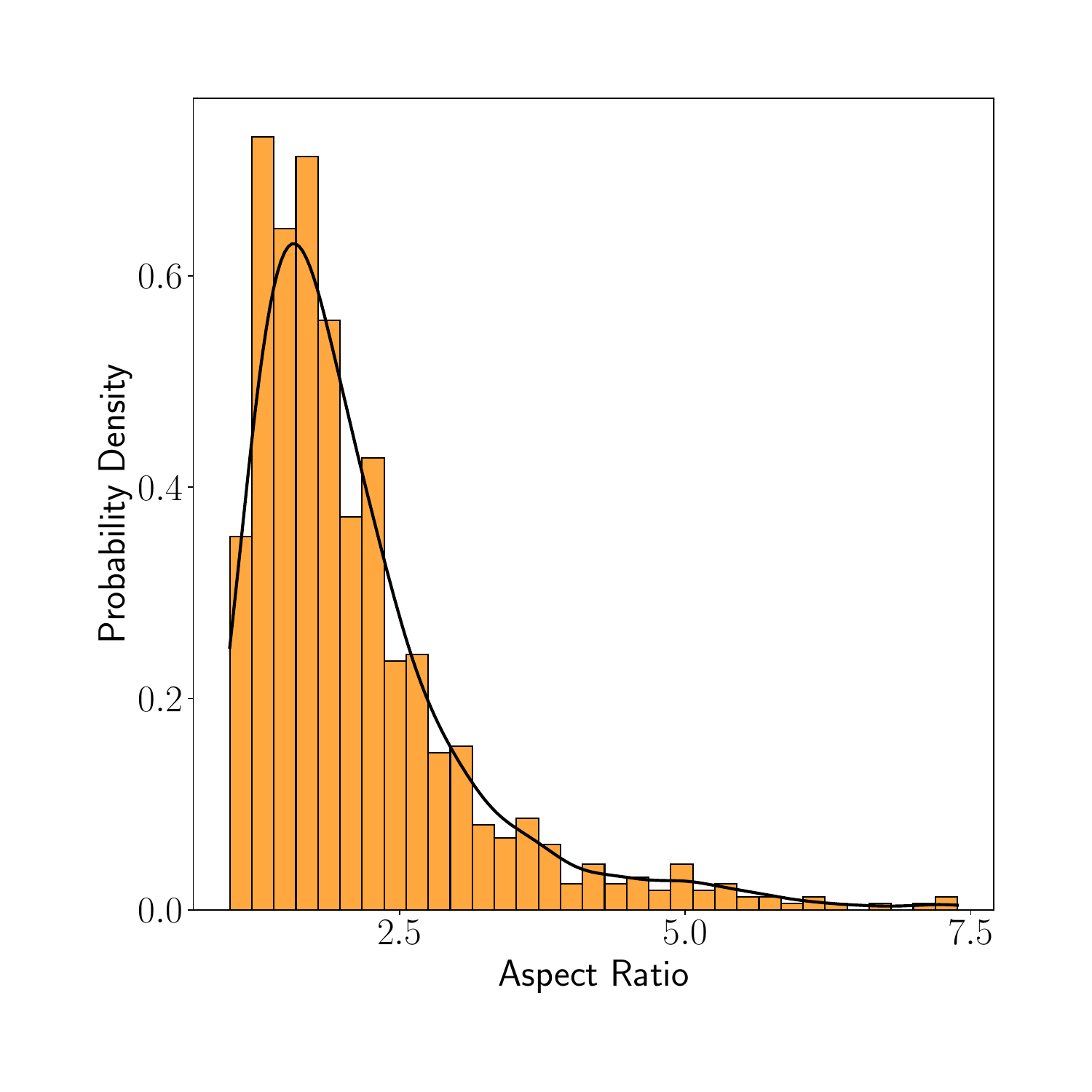}
        \caption{Crystal major axis by minor axis length ratio}
        \label{fig:CrystalAspectRatioHistogram}
    \end{subfigure}
    \hspace{0.05\linewidth}
    \begin{subfigure}[b]{0.40\linewidth} \centering
        \centering
        \begin{tikzpicture}
            \begin{axis}[
            xmin = 0,
            xmax = 45,
            ymin = 0,
            ymax = 45,
            domain=0:38,
            axis lines = middle,
            width=\linewidth,
            height=1.0\linewidth,
            x label style={at={(axis description cs:0.5,-0.1)},anchor=north},
            y label style={at={(axis description cs:-0.1,.5)},rotate=90,anchor=south},
            xlabel = {\scriptsize Major axis $(nm)$}, 
            ylabel = {\scriptsize Minor axis $(nm)$}]
            \addplot [scatter, only marks, color=orange, mark=o] table [x index= {6}, y index= {7}, col sep=comma] {data/ds_area_filtered_overall.csv};
            \addplot {x};
            \end{axis}
        \end{tikzpicture}
        \caption{Crystal major axis length vs. minor axis length}
        \label{fig:CrystalMajorVsMinorScatter}
    \end{subfigure}
    \caption{Individual crystal analysis. The properties of the individually detected crystals are plotted: (a) Crystal area, (b) FFT evaluated d-spacing, (c) Angle difference between crystal's pattern and its major axis, (d) Crystal aspect ratio, and (e) crystal axis lengths. (a) to (d) are histograms with kernel density estimate, and (e) is a scatter plot from analysis of the entire dataset.} 
    \label{fig:1p9_0p7_DSpacingPlots}
\end{figure}


Using the parameter set listed in \cref{Appendix}{appendix:OptimumParameters}, we applied \GRATEVtwo{} to detect and analyze crystalline domains within the HRTEM images of PCDTBT. The segmentation results for two images are presented in \figref{fig:HRTEMDetectionResults}, where the identified crystals are highlighted. Each detected crystal is surrounded by a convex hull boundary, with a shaded region representing a more precise delineation of the crystallite. A line at the crystal centroid indicates its orientation. The extracted properties for these detected crystals are summarized in \tabref{table:HRTEMDetectionResults}, including their centroid coordinates, area, orientation angle, d-spacing, major and minor axis lengths, and axis angle. Additionally, the correlation measurements between pairs of crystals are provided in \tabref{table:HRTEMCrystalCorrelation}, featuring metrics such as metric distance, direct distance, and relative angle.

The detection of these crystalline domains allows for a comprehensive analysis of the microstructural features of PCDTBT. The d-spacing values shown in \figref{fig:CrystalDSpacingHistogram} range from approximately 1.1nm to 2.9nm, which are consistent with the expected lattice spacings associated with PCDTBT's crystalline structures \citep{lu2012bilayer}. Variations in d-spacing among the detected crystals may be attributed to differences in crystallite orientation, strain within the material, or inherent structural disorder due to the semi-crystalline nature of PCDTBT.


The areas of the detected crystals vary significantly as shown in \figref{fig:CrystalAreaHistogram}, with values ranging from approximately 14.89nm$^2$ to 2307.18nm$^2$. Larger crystal areas may correlate with improved charge transport properties, as larger crystalline domains can facilitate more efficient charge carrier mobility along the polymer chains \citep{Park2020HighEfficiencyOPV}. The aspect ratios, derived from the major and minor axis lengths, provide insights into the shapes of the crystals. Higher aspect ratios indicate elongated, rod-like crystals, while lower aspect ratios suggest more equiaxed or spherical shapes. The diversity in crystal shapes and sizes can impact the overall morphology and performance of the polymer in electronic applications.


\figref{fig:1p9_0p7_DSpacingPlots} also presents additional statistical analyses of individual crystal properties across the dataset, including histograms of crystal areas, d-spacings, orientation angles, and shape descriptors like aspect ratio. These plots allow us to identify prevalent features and distributions within the material. For example, the histogram of d-spacings may show a peak around a specific value, indicating a dominant crystalline phase or preferred stacking distance within the polymer chains.


The comprehensive analysis provided by \GRATEVtwo{} enables us to establish quantitative relationships between microstructural features and potential material performance. By systematically characterizing the size, shape, orientation, and spatial distribution of crystalline domains, we can correlate these features with electronic properties measured in devices. This level of detailed microstructural understanding is essential for guiding the design and processing of conjugated polymers to achieve optimal performance in organic electronic applications.


For a more extensive and detailed exploration of this PCDTBT dataset, including insights into intercrystalline correlations and preferred crystallographic alignments, readers are referred to the work of our collaborators in \citep{fair2025automated}. Their study leverages automated HRTEM and the \GRATEVtwo{} image processing outputs to unravel how neighboring crystals preferentially align along certain lattice directions, likely reflecting underlying liquid crystalline order within the polymer. By combining the analysis presented here with their comprehensive assessment of orientation correlations and lattice parameters, one obtains a richer and more complete understanding of the polymer’s nanoscale structure and its implications for organic electronics.



\subsection{Timing Statistics}
\label{sec:results_timing}

\begin{figure}[b!]
    \centering
    \begin{tikzpicture}
        \begin{axis}[
            title={Timing Statistics},
            xlabel={Time (s)},
            xbar,
            xmin=0,
            y axis line style={opacity=0},
            axis x line=bottom,
            ytick=data,
            yticklabel style={align=right, font=\small},
            nodes near coords,
            nodes near coords align={horizontal},
            enlarge y limits=0.15,
            symbolic y coords={
                {Skeletonization},
                {Breaking Branches},
                {Blurring + Hist Eq + Thresholding},
                {Ellipse Construction},
                {d-Spacing Evaluation},
                {Segmentation},
                {Uniform Breaking},
                {Connected Component},
                {Adjacency Matrix}
            },
            y dir=reverse,
            bar width=15pt,
            width=12cm,
            height=8cm,
            grid=both,
            minor grid style={gray!25},
            major grid style={gray!25},
        ]
            \addplot[fill=blue!50] coordinates {
                (4.656,{Skeletonization})
                (0.912,{Breaking Branches})
                (0.238,{Blurring + Hist Eq + Thresholding})
                (0.207,{Ellipse Construction})
                (0.181,{d-Spacing Evaluation})
                (0.121,{Segmentation})
                (0.095,{Uniform Breaking})
                (0.049,{Connected Component})
                (0.043,{Adjacency Matrix})
            };
        \end{axis}
    \end{tikzpicture}
    \caption{Time taken by each step in the algorithm for the analysis of 1.9\,nm d-spacing crystals in a single image using a single core. The total processing time is approximately 6.52 seconds.}
    \label{fig:TimingBarChart}
\end{figure}
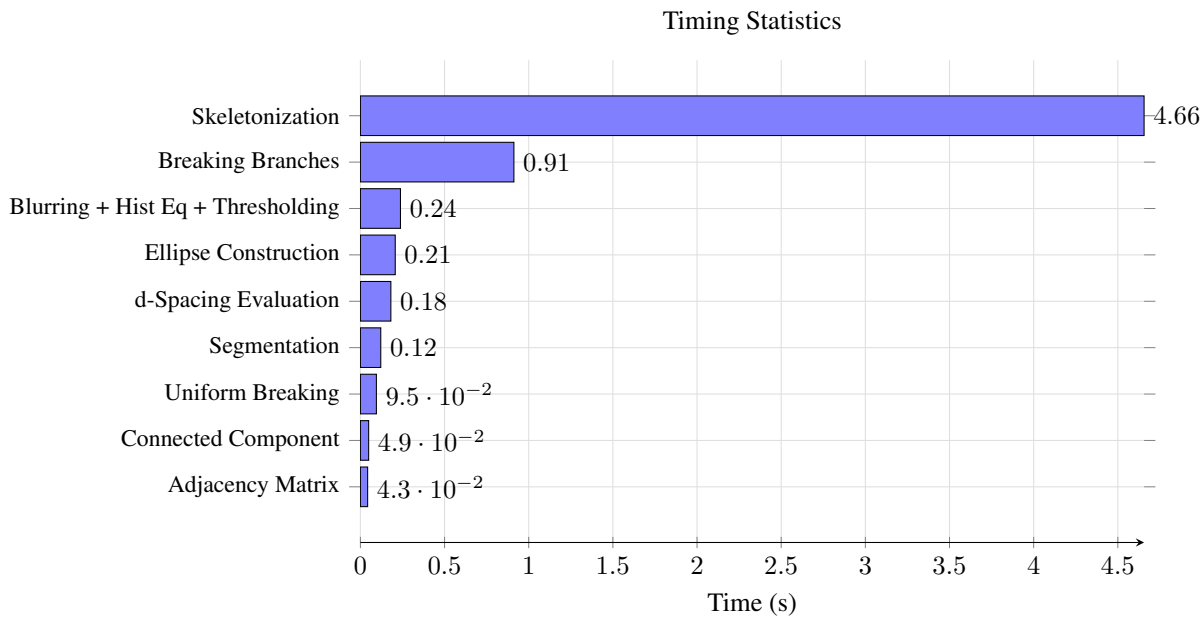

The algorithm was executed on a computer equipped with a 96-core AMD EPYC 9654 CPU @ 3.7\,GHz running Linux OS. The total time for processing a single high-resolution transmission electron microscopy (HRTEM) image with 1.9\,nm d-spacing crystals is approximately 6.52\,seconds when utilizing a single core. The timing consumption by various parts of the algorithm is presented in \figref{fig:TimingBarChart}. The most time-consuming steps are skeletonization (approximately 4.66\,seconds), followed by breaking branches and the preprocessing steps (blurring, histogram equalization, and thresholding). Utilizing all 96 cores, we processed an entire dataset of 637 images in 284\,seconds (approximately 4~minutes and 44~seconds), reducing the per-image processing time to just 0.44\,seconds. This significant improvement demonstrates the scalability and efficiency of our algorithm when parallelized across multiple cores.

The performance enhancements of our algorithm are attributed to the use of optimized libraries and functions that efficiently handle computationally intensive tasks. Specifically, we employed:
\begin{enumerate}
    \item The \texttt{skeletonize} function from \texttt{skimage.morphology} \citep{van2014scikit} for efficient skeletonization.
    \item OpenCV \citep{bradski2000opencv} functions \texttt{equalizeHist} and \texttt{threshold} (\texttt{cv2.equalizeHist}, \texttt{cv2.threshold}) for fast histogram equalization and image thresholding.
    \item Morphological operations such as closing and opening using \texttt{cv2.morphologyEx} from OpenCV.
    \item The \texttt{skeleton\_to\_csgraph} function from the \texttt{skan} \citep{nunez2018new} library for converting skeleton images to graph representations.
    \item The \texttt{label} function from \texttt{skimage.measure} \citep{van2014scikit} for rapid image segmentation.
    \item Sparse graph structures and connected component analysis using \texttt{csr\allowbreak\_matrix} and \texttt{connected\_compon\-ents} from \texttt{scipy.sparse.csgraph} \citep{virtanen2020scipy} for efficient graph construction and evaluation.
    \item Fast Fourier transforms and other numerical operations using NumPy functions such as \texttt{np.fft} \citep{harris2020array}.
    \item The \texttt{alphashape} library to create shrink-wraps around point clouds.
\end{enumerate}

By utilizing these optimized libraries, we minimized computational overhead and maximized the efficiency of each processing step. The skeletonization step, although still the most time-consuming, benefits significantly from the optimized implementation in \texttt{skimage}~\citep{van2014scikit}. Similarly, the use of sparse matrices and efficient graph algorithms from \texttt{scipy.sparse.csgraph}~\citep{virtanen2020scipy} greatly accelerates the analysis of the skeleton's connectivity.

Our analysis of the timing statistics reveals that the skeletonization step accounts for approximately 71\% of the total processing time for a single image when executed on a single core. This indicates that skeletonization is a major computational bottleneck in the algorithm. However, due to the parallel nature of image processing tasks, distributing the workload across multiple cores significantly mitigates this bottleneck. By processing images concurrently, we effectively utilize the available computational resources, resulting in a substantial reduction in total processing time.

Furthermore, the efficient handling of large data structures, such as sparse matrices in graph construction and connected component analysis, contributes to the algorithm's scalability. The use of optimized libraries ensures that even computationally intensive tasks are executed as efficiently as possible. This optimization is crucial when dealing with large datasets, as it enables rapid analysis without compromising accuracy or resolution.

The combination of algorithmic efficiency, optimized library functions, and effective parallelization allows our method to achieve high performance in processing and analyzing large volumes of HRTEM images. This capability is essential for applications requiring rapid data analysis and real-time feedback in materials science and related fields.

\subsection{ Data Sufficiency Analysis of PCDTBT Crystals in HRTEM Images}
\label{sec:results_dataSufficiency}

\begin{figure}[t!]
    \centering
    \begin{subfigure}[b]{0.24\linewidth}
        \centering
        \includegraphics[width=\linewidth,trim={0.6in 0.6in 0.6in 0.6in}, clip]{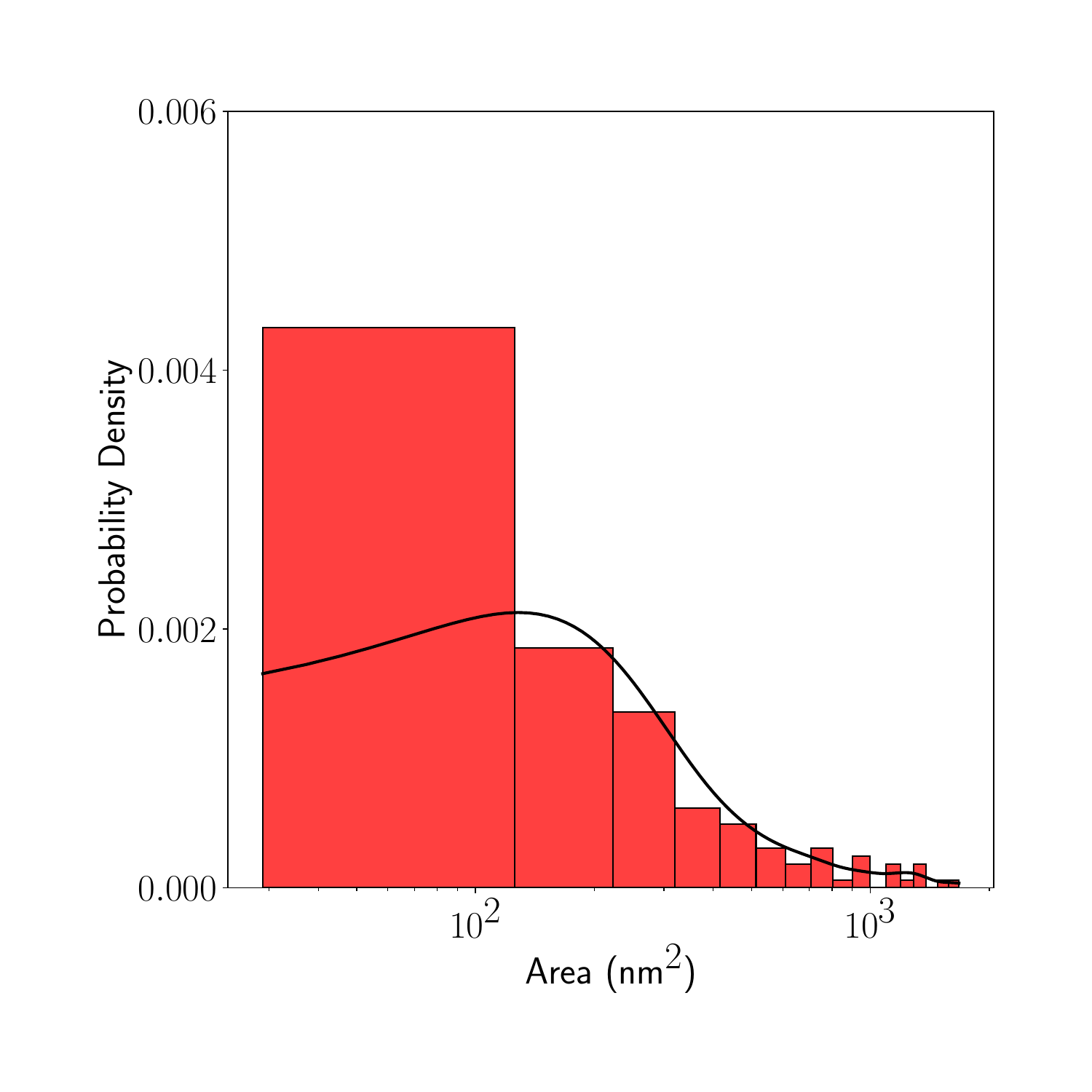}
        \caption{168 Crystals (20\%)}
        \label{fig:20per_histo_DataSuff}    
    \end{subfigure} 
    \begin{subfigure}[b]{0.24\linewidth}
        \centering
        \includegraphics[width=\linewidth,trim={0.6in 0.6in 0.6in 0.6in}, clip]{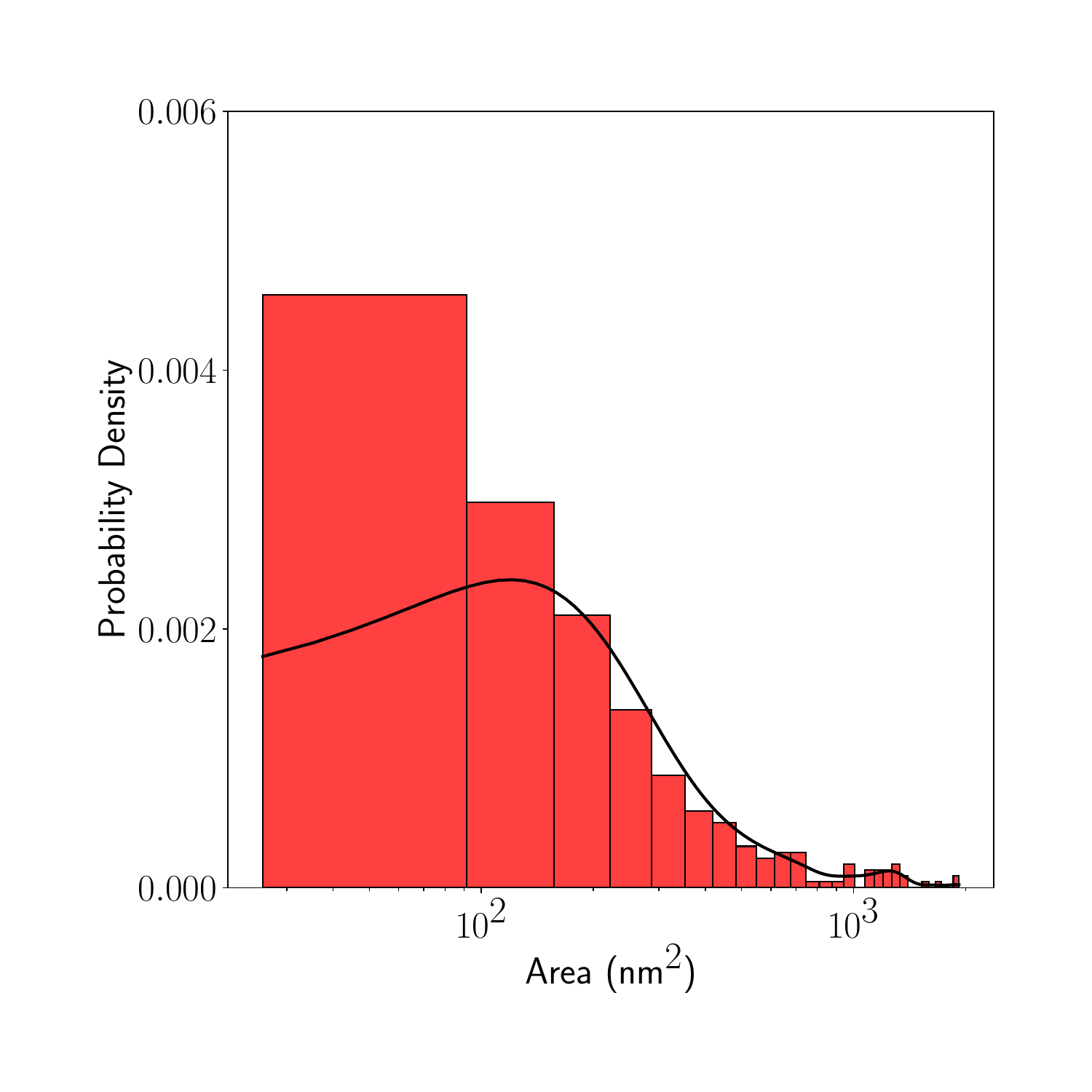}
        \caption{336 Crystals (40\%)}
        \label{fig:40per_histo_DataSuff}
    \end{subfigure}
    \centering
    \begin{subfigure}[b]{0.24\linewidth}
        \centering
        \includegraphics[width=\linewidth,trim={0.6in 0.6in 0.6in 0.6in}, clip]{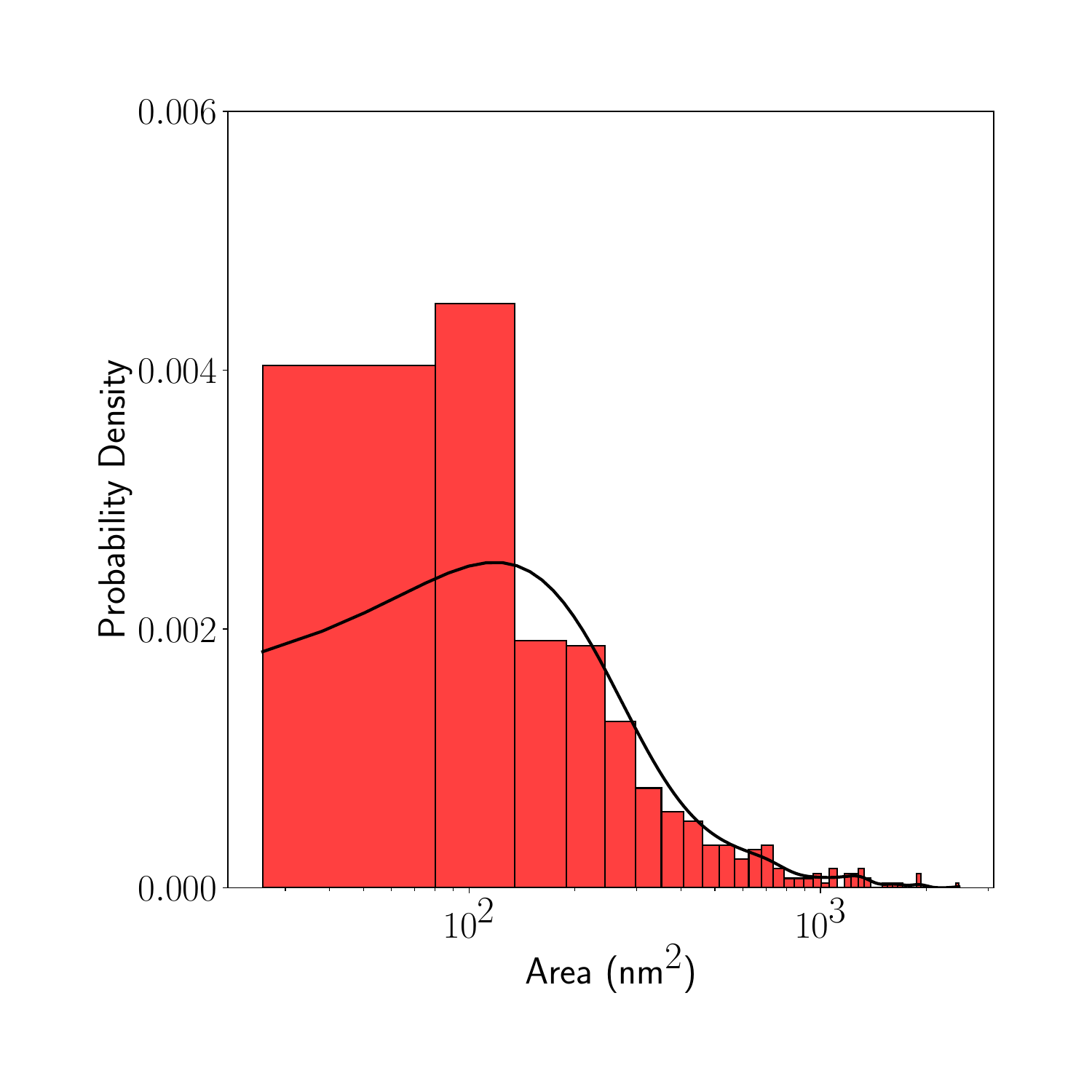}
        \caption{504 Crystals (60\%)}
        \label{fig:60per_histo_DataSuff}
    \end{subfigure}
    \begin{subfigure}[b]{0.24\linewidth}
        \centering
        \includegraphics[width=\linewidth,trim={0.6in 0.6in 0.6in 0.6in}, clip]{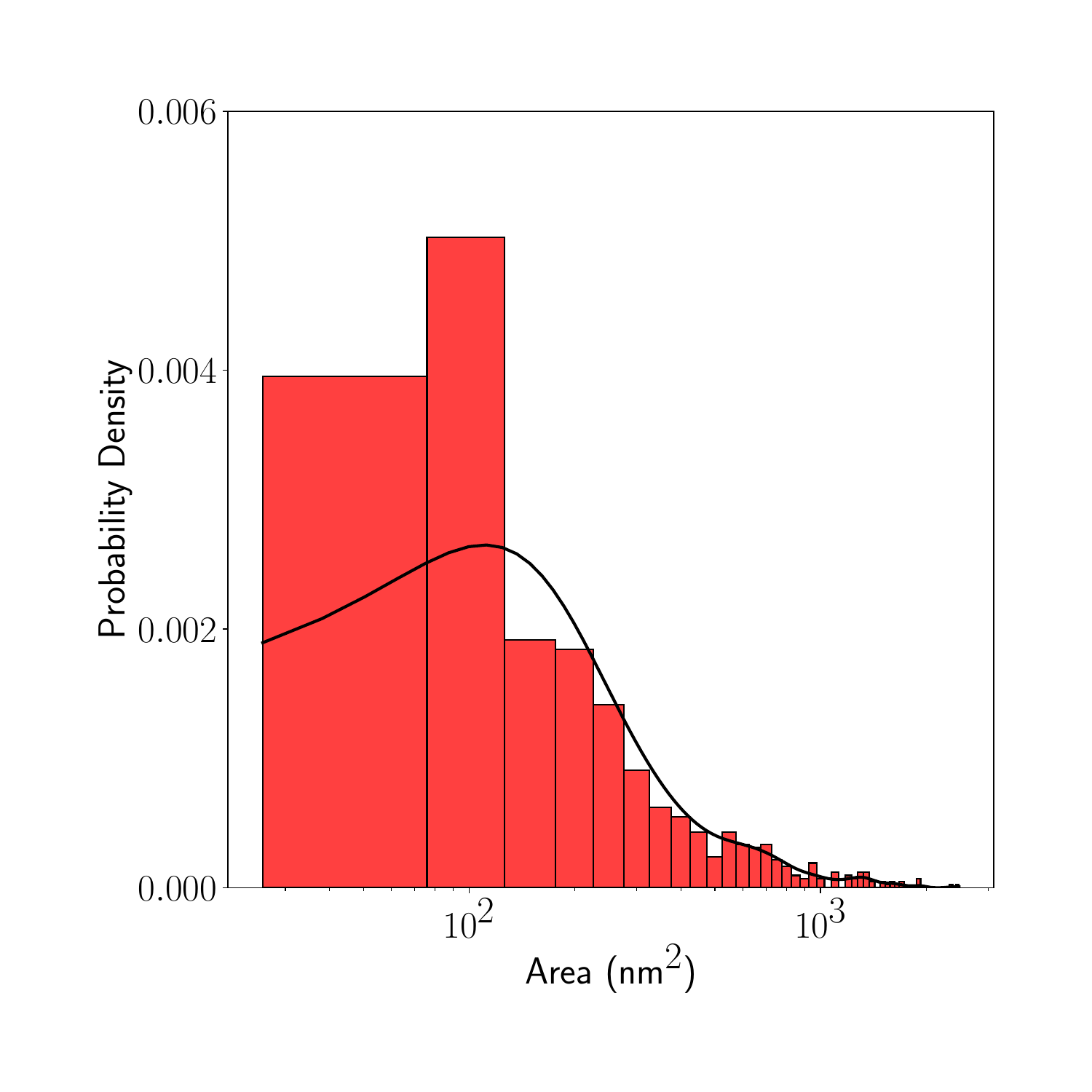}
        \caption{837 Crystals (100\%)}
        \label{fig:100per_histo_DataSuff}
    \end{subfigure}

    \begin{subfigure}[b]{0.7\linewidth}
        \centering
        \begin{tikzpicture}
        \begin{axis}[
            width=0.99\linewidth,
            height=0.6\linewidth,
            xlabel={Crystal Count},
            ylabel={Avg. Wasserstein Distance},
            xmin=0, xmax=850,
            ymin=0, ymax=55,
            xtick={0,100, 200, 300, 400, 500, 600, 700, 800},
            ytick={0,5,10,15,20,25,30,35,40,45,50, 55},
            ymajorgrids=true,
            grid style=dashed,
        ]
            \addplot[
                color=blue,
            ]
            table[
                x index=0,
                y index=1,
                col sep=comma
            ] {data/wassDist_currPrev_10.csv};
            \addlegendentry{\footnotesize Batch Size = 84 Crystals};
            
            \addplot[
                color=red
            ]
            table[
                x index=0,
                y index=1,
                col sep=comma
            ] {data/wassDist_currPrev_20.csv};
            \addlegendentry{\footnotesize Batch Size = 42 Crystals};
            
            \addplot[
                color=green,
            ]
            table[
                x index=0,
                y index=1,
                col sep=comma
            ] {data/wassDist_currPrev_40.csv};
            \addlegendentry{\footnotesize Batch Size = 21 Crystals};
            
            \addplot[
                color= orange,
            ]
            table[
                x index=0,
                y index=1,
                col sep=comma
            ] {data/wassDist_currPrev_80.csv};
            \addlegendentry{\footnotesize Batch Size = 10 Crystals};
    
        \end{axis}
        \end{tikzpicture}
        \caption{Wasserstein distance between consecutive batch increments of crystal data.}
        \label{fig:DSTrend}
    \end{subfigure}

    \caption{Data sufficiency evaluation for different crystal areas.}
    \label{fig:data_sufficiency}
\end{figure}
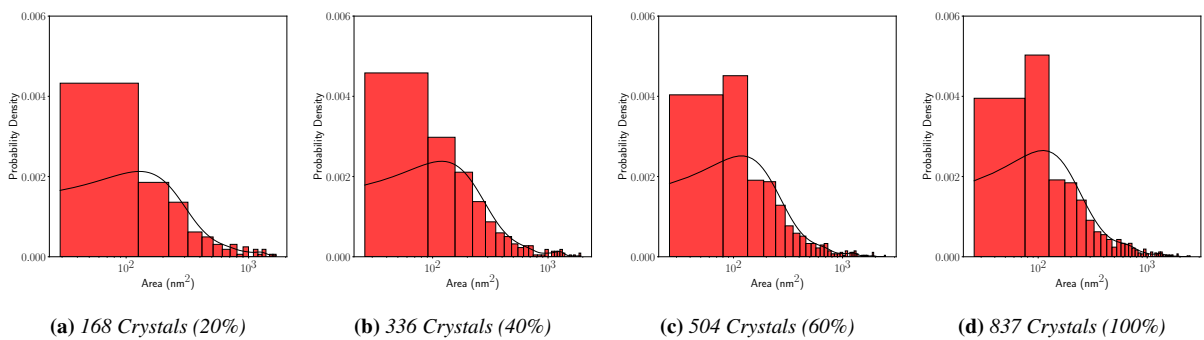
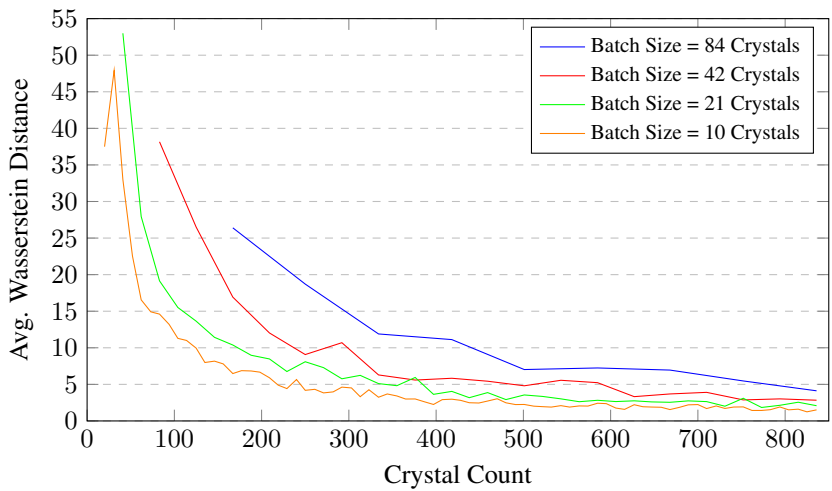

In our analysis of data sufficiency, we examine how the distribution of crystal areas reaches an asymptotic distribution as we incorporate more HRTEM images of PCDTBT. The complete dataset consists of 837 crystals from around 600 images. To evaluate how quickly the underlying distribution of crystal areas converges, we incrementally add data in fixed-size batches and compute the first-order Wasserstein distance \(W_1\) between distributions formed by consecutive increments.

Specifically, we consider four different batch sizes at which images (rather, crystals) are added to the dataset: 10, 21, 42, and 84 crystals. For each batch size scenario, we start from 0 crystals and progressively add data in increments equal to the batch size until we reach the full dataset size (of 837 crystals). After each increment, we compute the Wasserstein distance between the new distribution (including \(i \times \text{batchSize}\) crystals) and the previous distribution (with \((i-1) \times \text{batchSize}\) crystals). To obtain a more stable and reliable statistic, we repeat this evaluation 10 times at each increment, each time randomly sampling the \((i-1)\) batches from \(i\) batches of crystals and then average the resulting distances.

As shown in \figref{fig:data_sufficiency}, the histograms become smoother and more stable as the dataset grows, demonstrating that the distribution of crystal areas converges toward a steady form. Meanwhile, \figref{fig:DSTrend} shows how the averaged Wasserstein distance between consecutive increments decreases with additional data. Each curve represents a different batch size, revealing that the scale of incremental changes to the distribution depends on how many crystals are added at once.

Larger batch sizes (e.g., 84 crystals) introduce more data at each increment, leading to more pronounced changes in the distribution per step. Once enough large increments have been included, the distribution may show a sudden, relatively large drop in Wasserstein distance, then rapidly converge. In contrast, smaller batch sizes (e.g., 10 crystals) add data more gradually, producing smoother and more frequent updates. Each small increment makes a subtler change to the distribution, resulting in a more gradual and fine-grained trajectory toward convergence.

Because each batch size scenario scales the increments differently, the threshold for deciding when to stop data collection should also be scaled accordingly. For larger batch sizes, even a modest Wasserstein distance value (e.g., around 4 units for the 84-crystal batch) might indicate sufficient convergence, since one large increment can naturally shift the distribution more. For smaller batch sizes, where each increment is gentler (e.g., around 1.5 units difference for the 10-crystal batch), a lower threshold might be more appropriate, reflecting the finer control and resolution over the distribution’s shape. To illustrate this, \tabref{table:Wdist_batchComparison} shows the Wasserstein distance between the full dataset and a dataset that is one batch smaller, for each considered batch size scenario. These values provide concrete examples of how batch size influences the scale of change in the distribution.

\begin{table}[t!]
    \centering
    \caption{Wasserstein distance between the full dataset distribution and the distribution obtained by removing one batch of data from the full set, for various batch sizes.}
    \label{table:Wdist_batchComparison}
    \begin{tabular}{||c|c||}
        \hline
        \textbf{Batch Size} & \textbf{Wasserstein Distance} \\
        & \textbf{(Full vs. One Batch Less)} \\
        \hline\hline
        84 & 4.12 \\
        \hline
        42 & 2.83 \\
        \hline
        21 & 2.08 \\
        \hline
        10 & 1.51 \\
        \hline
    \end{tabular}
\end{table}

With this information, an experimentalist might set a higher threshold for larger batch sizes (e.g., around 4 units for an 84-crystal batch) and a lower threshold for smaller batch sizes (e.g., closer to 1.5 units for a 10-crystal batch). This ensures that the threshold for deciding when to cease data collection remains proportionate to the scale of changes induced by each incremental addition of data.

In practical terms, experimentalists can use insights from \figref{fig:DSTrend} and \tabref{table:Wdist_batchComparison} to tailor their data collection strategy. By selecting an appropriate batch size and a corresponding Wasserstein distance threshold, they can decide when further data provides diminishing returns. If quick feedback and high resolution of distribution changes are desired, a smaller batch size and a lower threshold can be chosen. If time or resources are limited, larger batch sizes and a slightly higher threshold may be more suitable. In either case, once the averaged Wasserstein distance consistently falls below the chosen threshold, the experimentalist can confidently cease data collection, knowing the distribution is sufficiently representative. This approach transforms data sufficiency from a guesswork exercise into a clear, data-driven criterion that guides experimental resource allocation and ensures that the resulting dataset meets the necessary statistical rigor. An experimentalist might:

\begin{itemize}
    \item \textbf{Begin Data Collection in Batches}: They determine a batch size (e.g., 10 crystals per batch) and start collecting HRTEM images in increments of that batch size.
    \item  \textbf{Periodic Assessment}: After each new batch of data, they compute the Wasserstein distance between the current and previous distributions of crystal areas. This computation can be done after every increment, providing immediate feedback on the impact of newly acquired data.
    \item \textbf{Decision Point}: If, after a certain number of increments, the averaged Wasserstein distance consistently falls below the established threshold (e.g., 1.5 units), the experimentalist has quantitative evidence that adding more data is unlikely to yield new insights into the crystal area distribution.
    \item \textbf{Stopping Data Collection}: With this statistical criterion, the experimentalist can confidently stop collecting further HRTEM images, reallocating their time and resources. This prevents unnecessary prolonged imaging campaigns.
\end{itemize}

\section{Conclusions}
\label{sec:conclusions}

In this work, we have developed and presented \GRATEVtwo{}, an open-source computational framework for the automated analysis of high-resolution transmission electron microscopy (HRTEM) images, specifically focusing on complex microstructures in conjugated polymers like PCDTBT. By leveraging fast, automated image processing algorithms augmented with Gaussian process optimization, \GRATEVtwo{} significantly reduces the need for manual selection of parameters and tuning, enhancing both reproducibility and user accessibility. The integration of a Wasserstein distance-based stopping criterion within \GRATEVtwo{} provides a quantitative method for optimizing data collection, ensuring efficient use of transmission electron microscopy (TEM) resources without compromising data quality.

\GRATEVtwo{}'s compatibility with HPC environments allows for efficient, large-scale data processing at near real-time speeds, making it suitable for high-throughput applications in materials science. By successfully applying \GRATEVtwo{} to a substantial PCDTBT dataset, we demonstrated its efficacy in rapidly extracting critical structural features such as d-spacing, orientation, and shape metrics. This capability is particularly valuable for advancing research in organic electronics, where precise nanoscale characterization is essential for optimizing material properties.

Overall, \GRATEVtwo{} addresses key limitations of existing HRTEM analysis methods by providing a fast, adaptable, and user-friendly tool that enhances the efficiency and reliability of microstructural characterization. By making \GRATEVtwo{} open-source, we aim to facilitate its adoption and further development by the research community. Future work could involve extending \GRATEVtwo{} to other material systems and incorporating additional analytical capabilities, thereby broadening its applicability and impact in the field of materials characterization.

\section{Methods}
\subsection{HRTEM Sample Preparation and Measurement Method}

PCDTBT was synthesized using previously published procedures for Suzuki polycondensation in a Schlenk reactor flask\citep{Xie-2018-Macromolecules, Xie-2017-Macromolecules}. Sigma-Aldrich supplied all reactants. Polymerization occurred between 9-(9-heptadecanyl)-9H-carbazole-2,7-diboronic acid bis(pinacol) ester and 4,7-bis(2-bromo-5-thienyl)-2,1,3-benzothiadiazole in toluene, with equimolar amounts of each monomer. All other synthesis and purification procedures remained unaltered compared to the cited sources. The synthesis product was characterized by H$^1$ nuclear magnetic resonance at 500 MHz\citep{fair2021molecular}. 

5 mg/mL solutions of PCDTBT and chlorobenzene (Sigma-Aldrich) were prepared in a nitrogen glovebox and mixed overnight on a hotplate at 45℃. Silicon wafers were sonicated in acetone for 20 minutes, then isopropanol for 20 minutes. The wafers then underwent 20 minutes of ultraviolet light ozonation. A PEDOT:PSS (Clevios P and H.C. Starck) and water solution were then spin-coated onto the clean substrates at 4000 RPM for 2 minutes. The samples were then brought into the glovebox, and the heated PCDTBT solution was spin-coated on top of the PEDOT:PSS layer at 800 RPM for 2 minutes. The sample was then cut into squares and floated off in the water, and samples were collected on copper TEM grids. Samples were left to dry overnight and then annealed inside a nitrogen glovebox.

High-resolution imaging experiments were conducted on the Titan Krios microscope at the Penn State Materials Characterization Laboratory. The accelerating voltage was 300 kV, and the detector was a Falcon 3EC direct electron detector in counted mode. Regions of interest were spaced 2.5 µm apart and visually inspected on the atlas image for tears and defects before acquisition. The spot size was set to 5, and autofocus was done at 300 kx magnification before being increased to 470 kx for acquiring a 2.5 second exposure. The microscope produced a 650 nm beam with a dose rate of 50 e/Å$^2$s.\par

\subsection{Quantities of Interest}
\label{QuantOfInterest}

The inputs to the algorithm are 1{)} the HRTEM image 2{)} Approx d-spacings of the crystals to detect 3{)} The resolution of the image 4{)} Process parameters. The algorithm outputs the segmentation result for each input and a CSV file containing the feature details as listed in \tabref{table:HRTEMDetectionResults} and \tabref{table:HRTEMCrystalCorrelation}.

The distance between the crystals are defined as below:
\begin{equation}
    D_\text{metric} = \frac{d_\text{center-center}}{r_1 + r_2}
    \label{eq:metricDistance_new}
\end{equation}
\begin{equation}
    D_\text{direct} = d_\text{center-center}
    \label{eq:directDistance_new}
\end{equation}

\begin{center}
    \renewcommand{\arraystretch}{1.5} 
    \begin{tabular}{>{\bfseries}l p{0.8\linewidth}} 
    \toprule
    \textbf{Symbol} & \textbf{Description} \\
    \midrule
    $D_\text{metric}$ & \textit{Metric distance}: A dimensionless measure of the relative separation between two crystals, calculated by dividing the center-to-center distance by the sum of the radii of the crystals. \\
    $D_\text{direct}$ & \textit{Direct distance (nm)}: The actual center-to-center distance between two crystals, measured in nanometers (nm). \\
    $d_\text{center-center}$ & \textit{Center-to-center crystal distance (nm)}: The straight-line distance between the centers of two crystals, measured in nanometers. \\
    $r_1, r_2$ & \textit{Radii of the crystals (nm)}: The radii of each crystal in nanometers, where $r_1$ represents the radius of the first crystal and $r_2$ represents the radius of the second. \\
    \bottomrule
    \end{tabular}
\end{center}

\subsection{Image Processing Algorithm}

\begin{figure}[ht]
    \centering
    \begin{tikzpicture}[node distance=1.0cm and 1.0cm, auto]

    \tikzstyle{input} = [rectangle, rounded corners, minimum width=2.5cm, minimum height=1cm, text centered, draw=black, fill=orange!30]
    \tikzstyle{process} = [rectangle, minimum width=3.5cm, minimum height=1cm, text centered, draw=black, fill=white, align=center]
    \tikzstyle{param} = [rectangle, draw=black, fill=blue!10, text width=3cm, text centered, font=\footnotesize, align=left]
    \tikzstyle{intermediate} = [rectangle, rounded corners, minimum width=0.6cm, minimum height=0.6cm, text centered, draw=black, fill=orange!30, align=center]
    \tikzstyle{output} = [rectangle, rounded corners, minimum width=2.5cm, minimum height=1cm, text centered, draw=black, fill=orange!30]
    \tikzstyle{arrow} = [thick,->,>=stealth]

    \node (inputimage) [process] {Input Image};
    \node (blurring) [process, below=of inputimage] {Blurring};
    \node (histeq) [process, below=of blurring] {Histogram\\ Equalization};
    \node (thresholding) [process, below=of histeq] {Thresholding};
    \node (closing) [process, below=of thresholding] {Closing and\\ Opening};
    \node (skeletonization) [process, below=of closing] {Skeletonization and\\ Branching};
    \node (segmentation) [process, below=of skeletonization] {Segmentation and\\ Filtering of Backbones};
    
    \node (breaking) [process, right=1.25cm of segmentation] {Breaking Backbones\\ in Uniform Size};
    \node (ellipse) [process, above=of breaking] {Ellipse\\ Construction};
    \node (ellipse_filter) [process, above=of ellipse] {Ellipse Aspect \\ Ratio Filtering};
    \node (adjacency) [process, above=of ellipse_filter] {Adjacency Matrix\\ Construction};
    \node (clustering) [process, above=of adjacency] {Cluster of\\ Adjacent Bones};
    \node (dspacing) [process, above=of clustering] {D-Spacing\\ Evaluation};
    \node (outputimage) [process, above=of dspacing] {Output Image};

    \draw [arrow] (inputimage) -- (blurring);
    \draw [arrow] (blurring) -- (histeq);
    \draw [arrow] (histeq) -- (thresholding);
    \draw [arrow] (thresholding) -- (closing);
    \draw [arrow] (closing) -- (skeletonization);
    \draw [arrow] (skeletonization) -- (segmentation);

    \draw [arrow] (segmentation) -- (breaking);

    \draw [arrow] (breaking) -- (ellipse);
    \draw [arrow] (ellipse) -- (ellipse_filter);
    \draw [arrow] (ellipse_filter) -- (adjacency);
    \draw [arrow] (adjacency) -- (clustering);
    \draw [arrow] (clustering) -- (dspacing);
    \draw [arrow] (dspacing) -- (outputimage);

    \node (param1) [param, left=0.75cm of blurring] {• Number of Iterations \\ • Kernel Size};
    \draw [dashed,->] (param1) -- (blurring);

    \node (param2) [param, left=0.75cm of closing] {• Kernel Size Closing \\ • Kernel Size Opening};
    \draw [dashed,->] (param2) -- (closing);

    \node (param3) [param, left=0.75cm of segmentation] {• Thresh. Pixel Length};
    \draw [dashed,->] (param3) -- (segmentation);

    \node (param4) [param, right=0.75cm of breaking] {• Breaking Length};
    \draw [dashed,->] (param4) -- (breaking);

    \node (param5) [param, right=0.75cm of ellipse_filter] {• Thresh. Aspect Ratio};
    \draw [dashed,->] (param5) -- (ellipse_filter);

    \node (param6) [param, right=0.75cm of adjacency] {• Adjacency Distance \\ • Adjacency Angle};
    \draw [dashed,->] (param6) -- (adjacency);

    \node (param7) [param, right=0.75cm of clustering] {• Thresh. Cluster Size \\ • Thresh. Area};
    \draw [dashed,->] (param7) -- (clustering);

    \node (param8) [param, right=0.75cm of dspacing] { • Band Filter Size \\ • Thresh. Power \\ Spectrum Peak };
    \draw [dashed,->] (param8) -- (dspacing);

    \node (start) [input, left=0.75cm of inputimage] {Algorithm Input};
    \draw [dashed] (start) -- (inputimage);

    \coordinate (mid_threshold_closingOpening) at ($(thresholding)!0.5!(closing)$);
    \node (intermediate1) [intermediate, left=2.50cm of mid_threshold_closingOpening] {1};
    \draw [dashed] (intermediate1) -- (mid_threshold_closingOpening);

    \coordinate (mid_skeleton_segment) at ($(skeletonization)!0.5!(segmentation)$);
    \node (intermediate2) [intermediate, left=2.50cm of mid_skeleton_segment] {2};
    \draw [dashed] (intermediate2) -- (mid_skeleton_segment);

    \coordinate (mid_breaking_ellipse) at ($(breaking)!0.5!(ellipse)$);
    \node (intermediate3) [intermediate, right=2.50cm of mid_breaking_ellipse] {3};
    \draw [dashed] (intermediate3) -- (mid_breaking_ellipse);

    \coordinate (mid_ellipFilter_adjacency) at ($(ellipse_filter)!0.5!(adjacency)$);
    \node (intermediate4) [intermediate, right=2.50cm of mid_ellipFilter_adjacency] {4};
    \draw [dashed] (intermediate4) -- (mid_ellipFilter_adjacency);

    \coordinate (mid_clustering_dspacing) at ($(clustering)!0.5!(dspacing)$);
    \node (intermediate5) [intermediate, right=2.50cm of mid_clustering_dspacing] {5};
    \draw [dashed] (intermediate5) -- (mid_clustering_dspacing);

    \node (output) [output, right=0.75cm of outputimage] {Algorithm Output};
    \draw [dashed] (outputimage) -- (output);

    \end{tikzpicture}
    \caption{Algorithm flowchart with parameters. Blue boxes are the process parameters and brown boxes are the intermediate steps shown in \figref{fig:vizAlgoOutputs}.}
    \label{fig:flowchartWithParameters}
\end{figure}
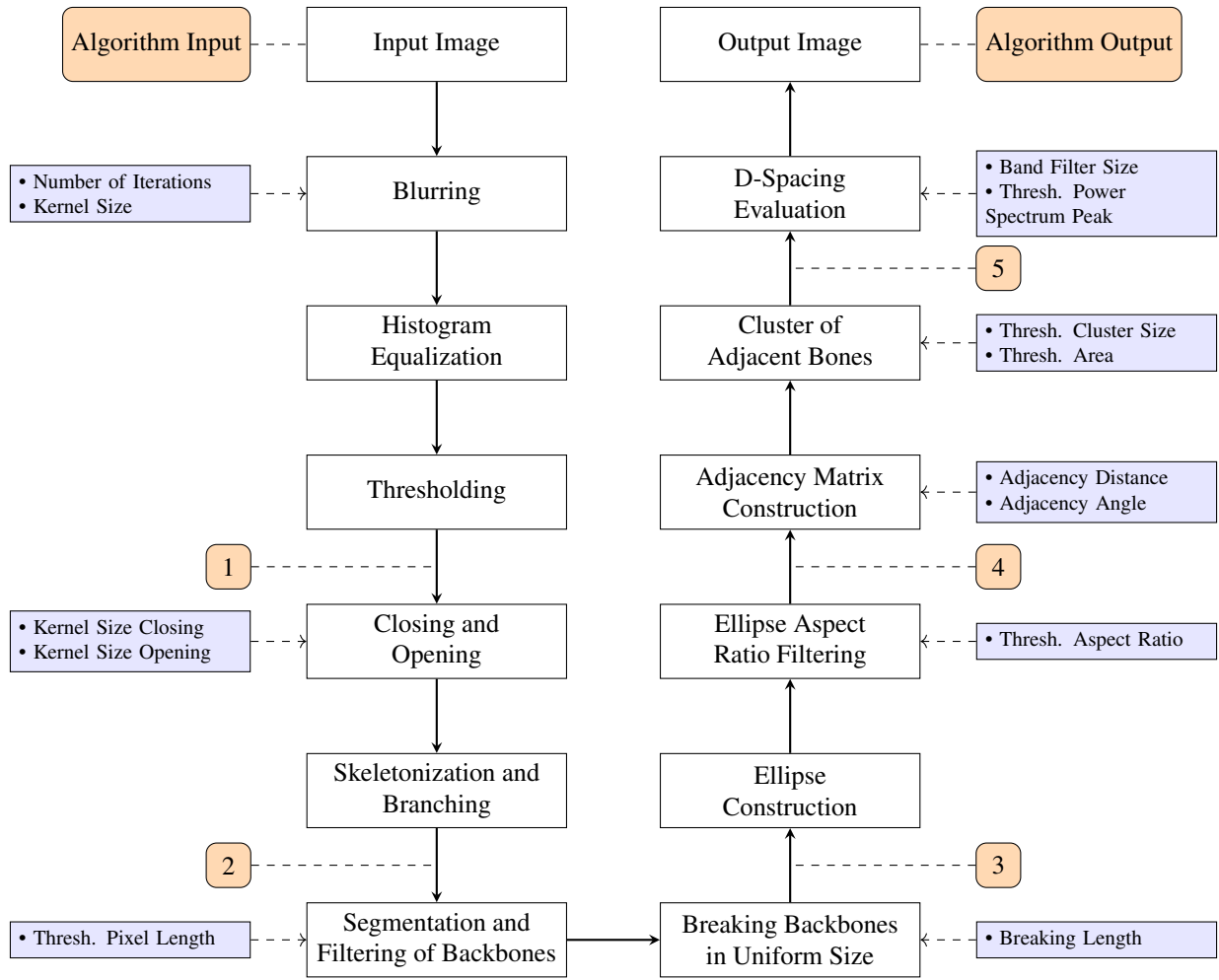

\begin{figure*}[t!]
    \centering
    \begin{subfigure}[b]{0.31\textwidth}
        \centering
        \includegraphics[width=\textwidth]{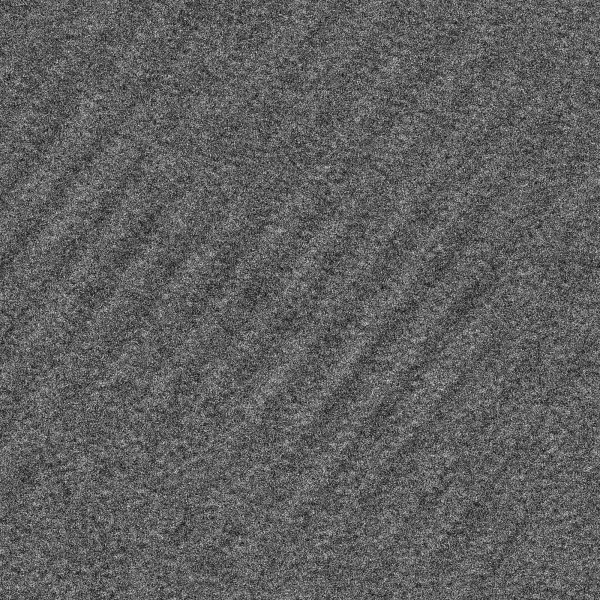}
        \caption{\textbf{Input:} Initial morphology}
        \label{fig:InitialMorphology}
    \end{subfigure}
    \hfill
    \begin{subfigure}[b]{0.31\textwidth}
        \centering
        \includegraphics[width=\textwidth]{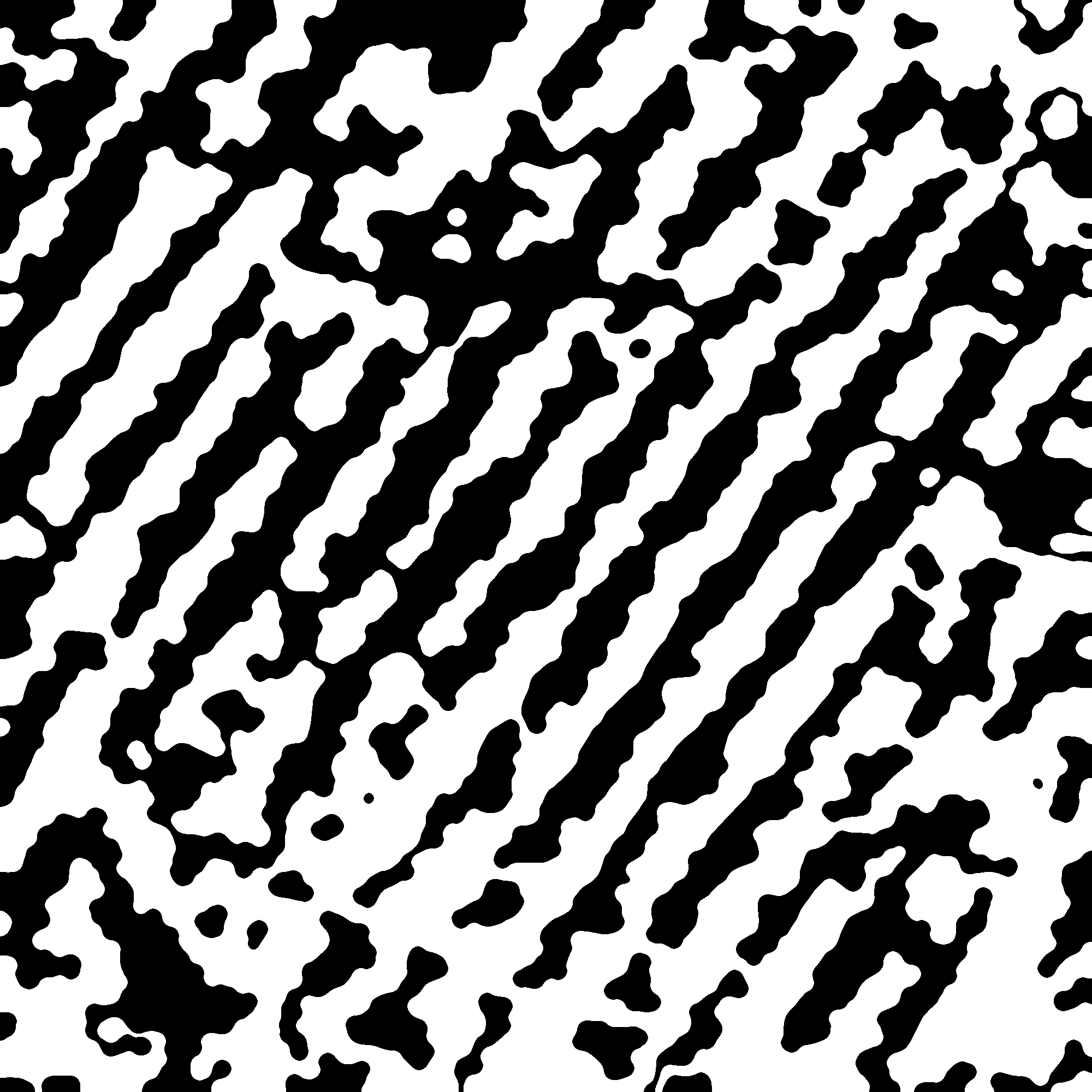}
        \caption{\textbf{1:} OTSU thresholding}
        \label{fig:OTSU_thresholding}
    \end{subfigure}
    \hfill
    \begin{subfigure}[b]{0.31\textwidth}
        \centering
        \includegraphics[width=\textwidth]{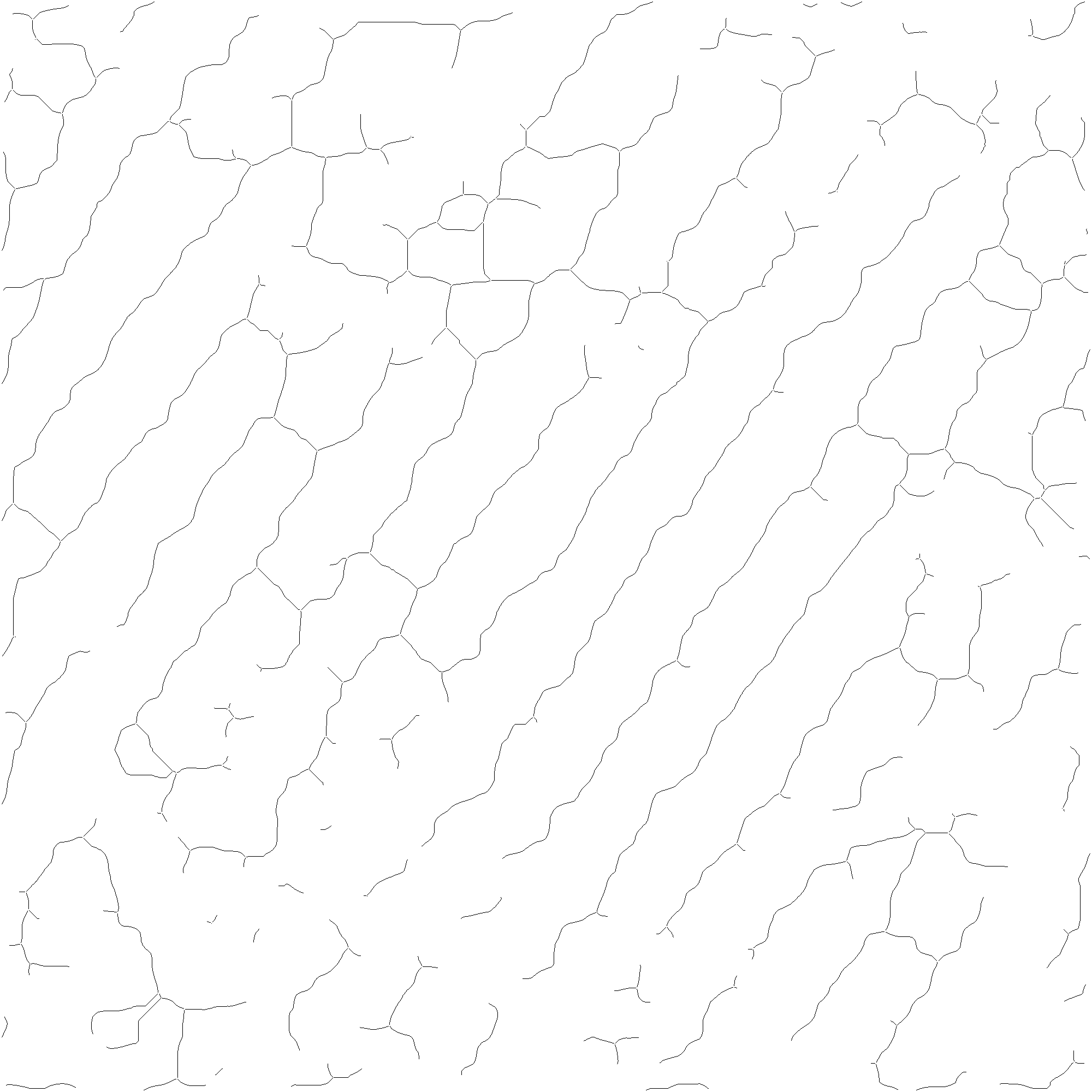}
        \caption{\textbf{2:} Skeletonization and branching}
        \label{fig:SkeletonizationAndBranching}
    \end{subfigure}
    \hfill
    \begin{subfigure}[b]{0.31\textwidth}
        \centering
        \includegraphics[width=\textwidth]{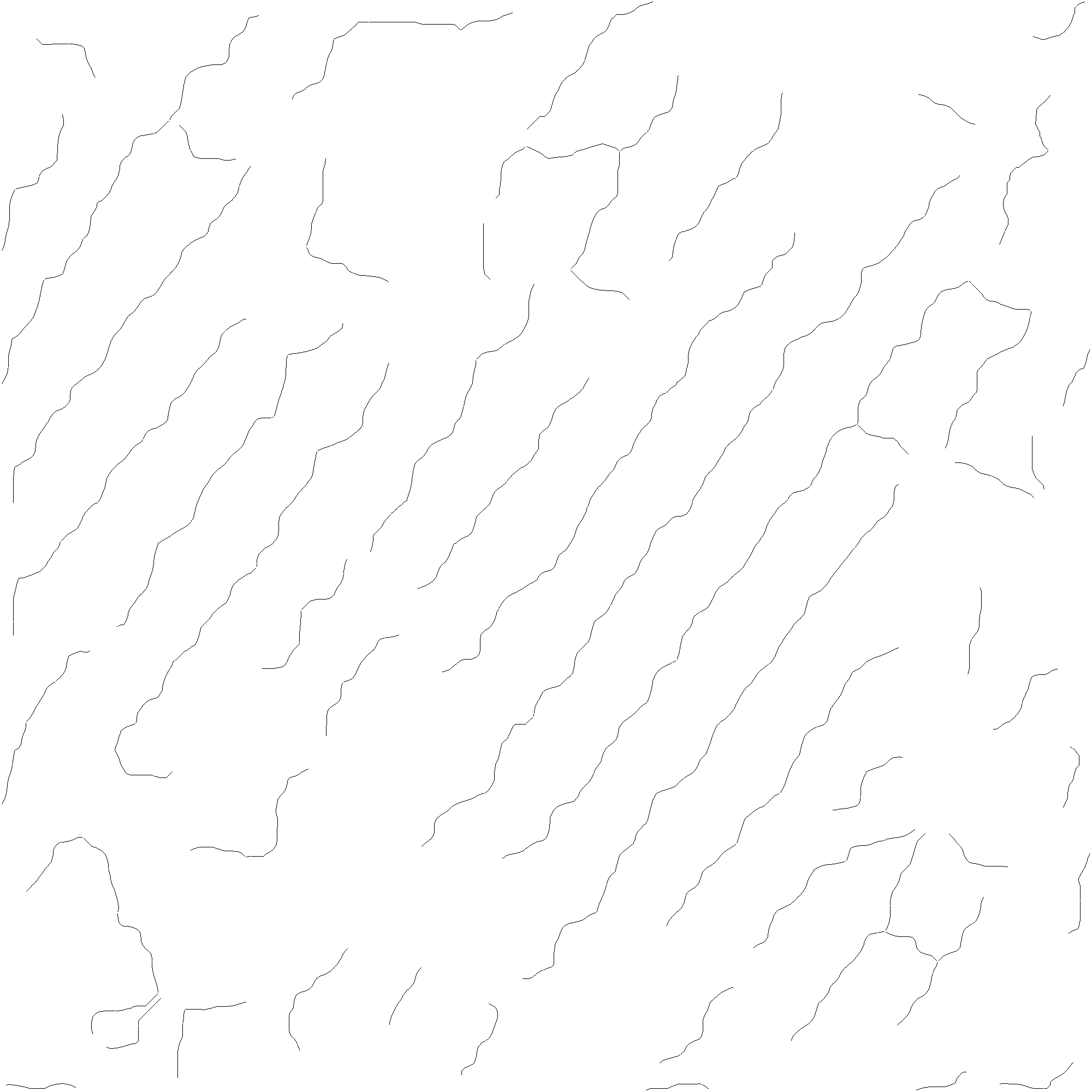}
        \caption{\textbf{3:} Filtering and bone division}
        \label{fig:filteringAndBoneDivision}
    \end{subfigure}
    \hfill
    \begin{subfigure}[b]{0.31\textwidth}
        \centering
        \includegraphics[width=\textwidth]{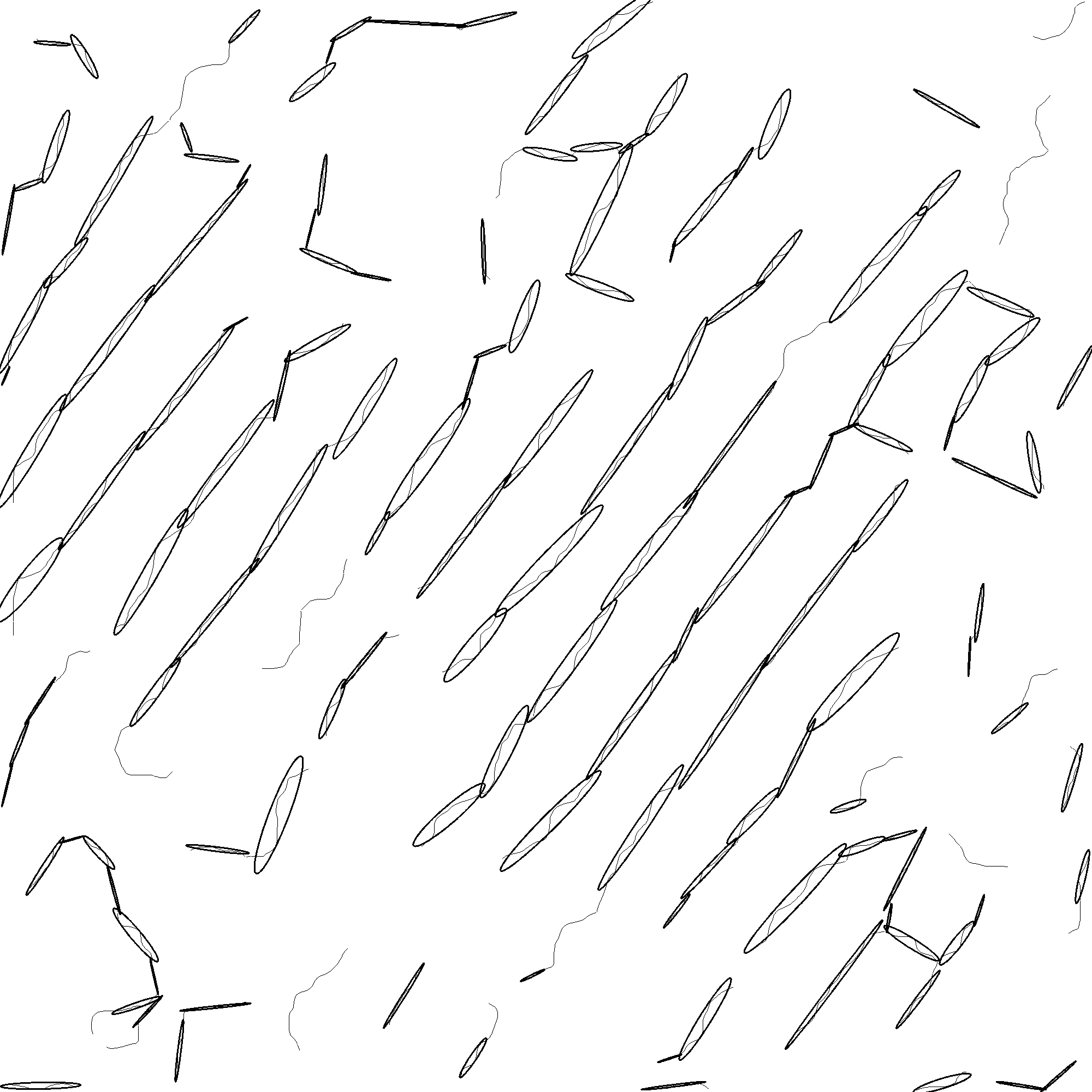}
        \caption{\textbf{4:} Aspect ratio filtering}
        \label{fig:aspectRatioFiltering}
    \end{subfigure}
    \hfill
    \begin{subfigure}[b]{0.31\textwidth}
        \centering
        \includegraphics[width=\textwidth]{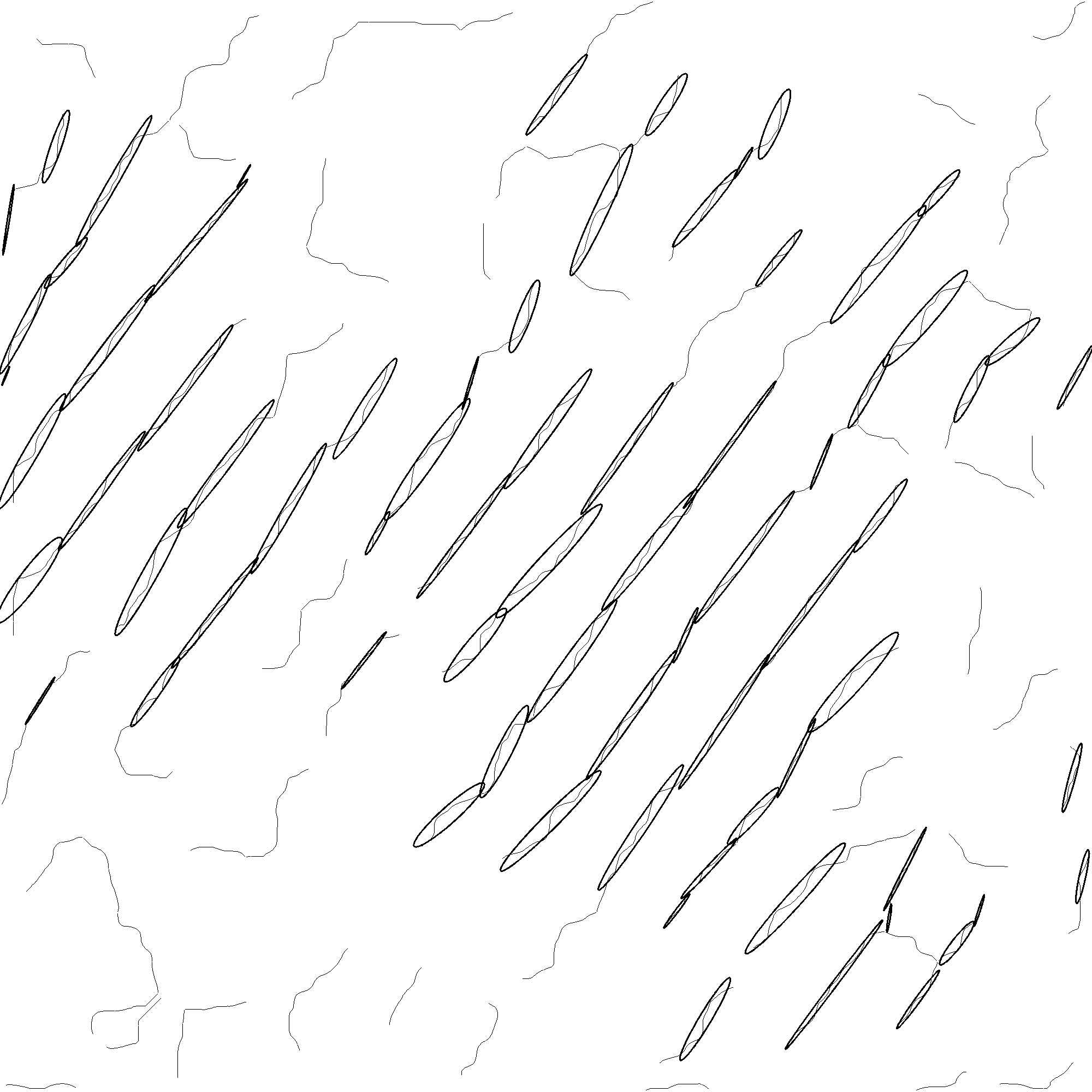}
        \caption{\textbf{5:} Clustering}
        \label{fig:clustering}
    \end{subfigure}
    \hfill
    \begin{subfigure}[b]{0.31\textwidth}
        \centering
        \includegraphics[width=\textwidth]{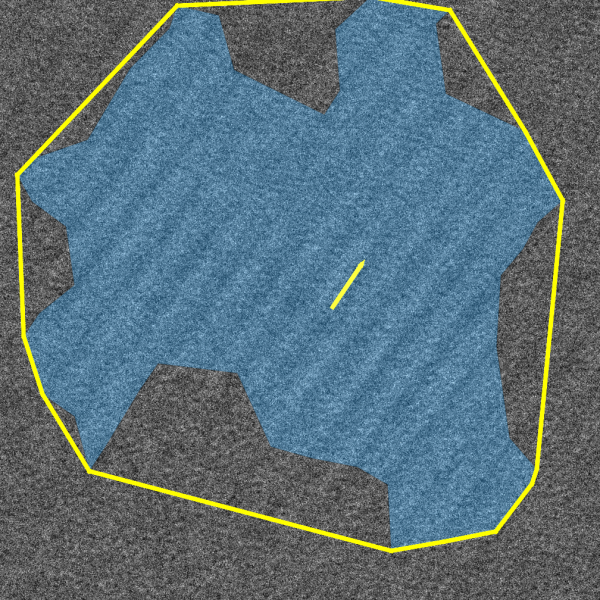}
        \caption{\textbf{Output:} Crystal region}
        \label{fig:crystalDetection}
    \end{subfigure}
    \caption{Visualization of intermediate and final algorithm outputs.}
    \label{fig:vizAlgoOutputs}
\end{figure*}

\begin{figure*}[t!]
    \centering
    \includegraphics[width=.8\columnwidth]{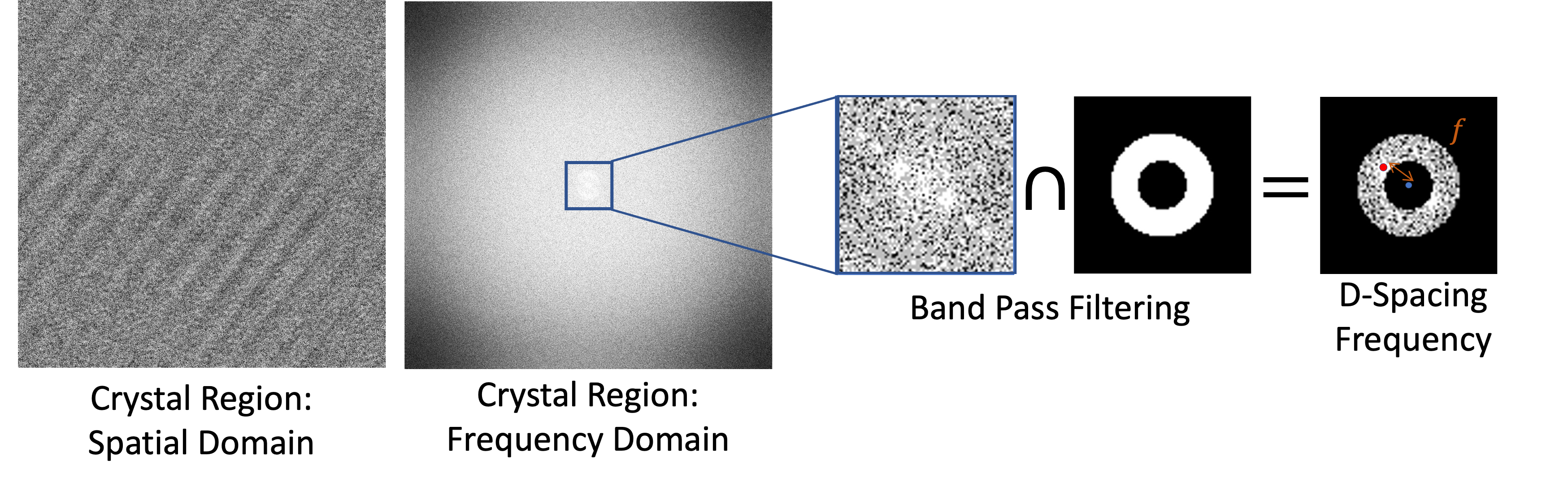}
    \caption{Evaluation of d-spacing for the crystal using the Fast Fourier Transform}
    \label{fig:fft_dspacing_evaluation}
\end{figure*}

\figref{fig:flowchartWithParameters} illustrates the flowchart of \GRATEVtwo{} algorithm along with the associated process parameters. The process parameters are indicated in blue boxes, while the intermediate steps corresponding to \figref{fig:vizAlgoOutputs} are marked with brown numeric annotations.

The algorithm parameters are classified into two categories: \textit{primary parameters} and \textit{secondary parameters}. The primary parameters are independent values provided by the user based on the dataset's imaging conditions and material properties, specifically the crystal d-spacing value and the image resolution. These parameters scale the key process parameters, enabling the algorithm to generalize across different imaging conditions and crystals of interest. They may vary significantly between datasets. Secondary parameters correspond to individual steps of the algorithm and are set to optimal values; they may require fine-tuning when changing datasets.

The algorithm is designed with several key objectives in mind: to generate a clear and concise representation of the essential information in the image; to enhance data quality by filtering out noise and irrelevant details; to convert the processed image data into a graph structure for advanced analysis; to use graph algorithms to identify clusters of polymer backbones forming crystallites; and to apply Fourier transform techniques to determine the d-spacing values.

Each step of the algorithm is detailed below, along with its associated parameters.
\begin{enumerate}
    \item Initially, \textbf{blurring} is applied to smooth the image and reduce sharpness, which helps in minimizing noise and preparing the image for subsequent processing. This operation is performed multiple times, with the number of blurring iterations specified as a parameter. Following blurring, \textbf{histogram equalization} is employed to increase the image contrast, making it sharper and enhancing the distinction between polymer chains and the background. Improved contrast facilitates more accurate thresholding in the subsequent step.
    \item The next step involves \textbf{image thresholding} using Otsu's method to create a binary representation of the image, where polymer chains appear as black regions and the background as white. The result of this step is shown in \figref{fig:OTSU_thresholding}. After thresholding, \textbf{morphological closing and opening} operations are performed to remove small black and white spots considered as noise. The kernel sizes for closing and opening are parameters that do not depend on the d-spacing and are set based on the noise characteristics of the image.
    \item Subsequently, we perform \textbf{skeletonization} and \textbf{branching}. Skeletonization reduces the polymer regions to single-pixel-wide lines, representing the skeleton of the polymer chains. Branching breaks the skeleton at junctions where three or more connections occur, resulting in branched skeletonized representations called backbones. The output of this step is depicted in \figref{fig:SkeletonizationAndBranching}.
    \item To refine the data, we proceed to \textbf{segment and filter the backbones}. Backbones shorter than a certain pixel length are considered noise and are filtered out; the threshold length is a parameter proportional to the d-spacing. The remaining backbones are segmented into uniform lengths, referred to as \textit{bones}, for consistent evaluation. The division size is also a parameter proportional to the d-spacing. The result after filtering and division is shown in \figref{fig:filteringAndBoneDivision}.
    \item For each bone, we perform \textbf{ellipse construction} by fitting an ellipse to its pixel locations using the \texttt{scikit\-image} library~\citep{vanDerWalt-2014-PeerJ}. This provides the major and minor axes of the bone, which are used for further analysis. Ellipses are preferred because they help filter out non-linear bones (curved structures not part of crystalline regions) and facilitate the creation of a graph representation where each ellipse serves as a node.
    \item To identify the crystalline regions, we apply \textbf{ellipse aspect ratio filtering}. Crystalline regions are composed of linear polymer backbones, so we filter out curved bones by setting a threshold on the ellipse aspect ratio (major axis length divided by minor axis length). Bones with aspect ratios above this threshold are retained for further analysis. This step's output is shown in \figref{fig:aspectRatioFiltering}.
    \item A \textbf{graph} is then constructed where each ellipse represents a node. An edge is created between two nodes if the distance between their centers is less than the adjacency distance parameter (proportional to the d-spacing) and the angle between their major axes is less than the adjacency angle parameter (also proportional to the d-spacing). The graph is stored as an adjacency matrix.
    
    Using this graph representation, we perform \textbf{clustering of adjacent bones} by identifying connected components using a depth-first search algorithm. Clusters with a node count below a certain threshold are considered noise and discarded. The remaining clusters represent detected crystalline regions. For each cluster, a convex hull and an alpha shape are computed to segment the crystal region. Properties such as area, centroid, major and minor axis lengths, and orientation are evaluated. Additionally, a line indicating the direction of the crystal pattern is plotted. The output after clustering is shown in \figref{fig:clustering}, and the final crystal detection is displayed in \figref{fig:crystalDetection}.

\end{enumerate}

Finally, we perform \textbf{d-spacing evaluation} for each detected crystal region. The largest possible square region within the crystal is selected, and a Fast Fourier Transform (FFT) is performed to transform the spatial domain into the frequency domain. Band-pass filtering is applied to remove frequencies outside the range of interest. The location of the peak frequency above a set threshold is used to calculate the exact d-spacing value and the orientation of the crystal pattern. The frequency threshold is a parameter. The evaluation of d-spacing using FFT is illustrated in \figref{fig:fft_dspacing_evaluation}. The algorithm outputs include visualizations such as convex hulls and alpha shapes of detected crystal regions overlaid on the original image, as well as quantitative data like area, centroid, major and minor axis lengths, orientation, and d-spacing values for each crystal. All features are saved in CSV files for further analysis. These outputs provide valuable insights into the material's microstructure and can aid in understanding material properties more effectively.

The algorithm is implemented in Python, utilizing the \texttt{scikit-image} library for image processing and SciPy for computational functions. A configuration file is used to input all relevant parameters, dataset paths, and result paths. The code is tested on Ubuntu Linux-based local systems and HPC servers. It supports batch processing and multiprocessing, allowing for efficient analysis of large datasets. Results are stored in a well-organized directory structure, with options to save intermediate outputs for debugging purposes. Comprehensive documentation is provided, detailing code usage and parameter settings. A summary of the primary and secondary parameters used in the algorithm is provided in \colref{Appendix}{appendix:OptimumParameters}. These parameters are optimized for detecting crystals in the HRTEM dataset of PCDTBT organic photovoltaic materials.

\subsection{Bayesian Optimization}
\label{sec:bayesian_optimization}

Optimizing hyperparameters in image processing algorithms is crucial for enhancing performance, especially in complex tasks like automated detection of crystalline domains in HRTEM images. Traditional methods of parameter tuning, such as grid search or manual adjustment, can be inefficient and may not guarantee optimal results due to the high dimensionality and computational expense. To address this challenge, we integrated Bayesian optimization into our framework, \GRATEVtwo{}, to systematically and efficiently identify the optimal set of parameters.

Bayesian optimization is a sequential design strategy for global optimization of black-box functions that are expensive to evaluate especially due to large hyperparameter space \citep{frazier2018tutorial}. It builds a probabilistic model of the objective function and uses it to select the most promising hyperparameters to evaluate next, balancing exploration and exploitation.

\subsubsection{Mathematical Formulation}

Let $\mathbf{x} \in \mathcal{X} \subseteq \mathbb{R}^d$ denote a vector of $d$ hyperparameters, and let $f: \mathcal{X} \rightarrow \mathbb{R}$ be the objective function that maps hyperparameters to a scalar performance metric. Our goal is to find the hyperparameters $\mathbf{x}^\ast$ that minimize the objective function:

\begin{equation}
\mathbf{x}^\ast = \arg \min_{\mathbf{x} \in \mathcal{X}} f(\mathbf{x}).
\end{equation}

In our case, $f(\mathbf{x})$ is the negative mean Intersection over Union (IoU) score between the detected crystalline regions and the ground truth annotations, as we aim to maximize the IoU.

\subsubsection{Gaussian Process Surrogate Model}

Bayesian optimization relies on a surrogate model to approximate the objective function. We use a Gaussian process (GP) prior \citep{rasmussen2003gaussian} over functions to model $f(\mathbf{x})$. A GP is defined by its mean function $m(\mathbf{x})$ and covariance function $k(\mathbf{x}, \mathbf{x}')$:

\begin{equation}
f(\mathbf{x}) \sim \mathcal{GP}\left( m(\mathbf{x}), k(\mathbf{x}, \mathbf{x}') \right).
\end{equation}

We assume a zero mean function $m(\mathbf{x}) = 0$ without loss of generality, and select a suitable covariance function, such as the squared exponential (radial basis function) kernel:

\begin{equation}
k(\mathbf{x}, \mathbf{x}') = \sigma_f^2 \exp\left( -\frac{1}{2} (\mathbf{x} - \mathbf{x}')^\top \mathbf{L}^{-1} (\mathbf{x} - \mathbf{x}') \right),
\end{equation}

where $\sigma_f^2$ is the signal variance and $\mathbf{L}$ is a diagonal matrix of length-scale parameters $\ell_i^2$, controlling the smoothness of the function along each dimension.

\subsubsection{Posterior Distribution}

Given a set of $n$ observed data points $\mathcal{D}_n = \{ (\mathbf{x}_i, y_i) \}_{i=1}^n$, where $y_i = f(\mathbf{x}_i) + \epsilon_i$ and $\epsilon_i \sim \mathcal{N}(0, \sigma_n^2)$ is Gaussian observation noise, the GP posterior predictive distribution at a new point $\mathbf{x}_\ast$ is given by:

\begin{align}
\mu_n(\mathbf{x}_\ast) &= \mathbf{k}_\ast^\top (\mathbf{K} + \sigma_n^2 \mathbf{I})^{-1} \mathbf{y}, \label{eq:posterior_mean} \\
\sigma_n^2(\mathbf{x}_\ast) &= k(\mathbf{x}_\ast, \mathbf{x}_\ast) - \mathbf{k}_\ast^\top (\mathbf{K} + \sigma_n^2 \mathbf{I})^{-1} \mathbf{k}_\ast, \label{eq:posterior_variance}
\end{align}

where $\mathbf{k}_\ast = [k(\mathbf{x}_1, \mathbf{x}_\ast), \dots, k(\mathbf{x}_n, \mathbf{x}_\ast)]^\top$, $\mathbf{K}$ is the $n \times n$ covariance matrix with $[\mathbf{K}]_{ij} = k(\mathbf{x}_i, \mathbf{x}_j)$, and $\mathbf{y} = [y_1, \dots, y_n]^\top$.

\subsubsection{Acquisition Function}

An acquisition function $\alpha(\mathbf{x}; \mathcal{D}_n)$ guides the selection of the next evaluation point by quantifying the utility of sampling at $\mathbf{x}$. We use the Expected Improvement (EI) acquisition function \citep{jones1998efficient}, defined as:

\begin{equation}
\alpha_{\text{EI}}(\mathbf{x}) = \mathbb{E}\left[ \max\left( 0, f_{\text{min}} - f(\mathbf{x}) \right) \right],
\end{equation}

where $f_{\text{min}}$ is the minimum observed value of the objective function. Under the GP posterior, the EI can be computed analytically:

\begin{equation}
\alpha_{\text{EI}}(\mathbf{x}) = (f_{\text{min}} - \mu_n(\mathbf{x})) \Phi\left( \frac{f_{\text{min}} - \mu_n(\mathbf{x})}{\sigma_n(\mathbf{x})} \right) + \sigma_n(\mathbf{x}) \phi\left( \frac{f_{\text{min}} - \mu_n(\mathbf{x})}{\sigma_n(\mathbf{x})} \right),
\end{equation}

where $\mu_n(\mathbf{x})$ and $\sigma_n(\mathbf{x})$ are the posterior mean and standard deviation from Eqs.~\eqref{eq:posterior_mean} and \eqref{eq:posterior_variance}, $\Phi(\cdot)$ is the standard normal cumulative distribution function, and $\phi(\cdot)$ is the standard normal probability density function.

\subsubsection{Optimization Loop}

\begin{figure}[ht]
    \centering
    \begin{tikzpicture}[node distance=1.25cm, auto]

    \tikzstyle{startstop} = [rectangle, rounded corners, minimum width=3cm, minimum height=0.75cm, text centered, draw=black, fill=gray!20]
    \tikzstyle{process} = [rectangle, minimum width=3cm, minimum height=0.75cm, text centered, draw=black, fill=orange!50]
    \tikzstyle{decision} = [diamond, aspect=3, minimum width=3cm, minimum height=1cm, text centered, draw=black, fill=blue!30]
    \tikzstyle{arrow} = [thick,->,>=stealth]

    \node (start) [startstop] {Start: Initialization};
    \node (process1) [process, below of=start] {Evaluate initial points $\{ \mathbf{x}_i \}_{i=1}^{n_0}$};
    \node (process2) [process, below of=process1] {Fit GP surrogate model};
    \node (process3) [process, below of=process2] {Maximize acquisition function:
    $\mathbf{x}_{n+1} = \arg \max_{\mathbf{x} \in \mathcal{X}} \alpha(\mathbf{x}; \mathcal{D}_n)$};
    \node (process4) [process, below of=process3] {Evaluate objective at $\mathbf{x}_{n+1}$ to get $y_{n+1}$};
    \node (process5) [process, below of=process4] {Augment data: $\mathcal{D}_{n+1} = \mathcal{D}_n \cup \{ (\mathbf{x}_{n+1}, y_{n+1}) \}$};
    \node (decision) [decision, below of=process5, yshift=-0.5cm] {Convergence criterion met?};
    \node (stop) [startstop, below of=decision, yshift=-0.5cm] {End};

    \draw [arrow] (start) -- (process1);
    \draw [arrow] (process1) -- (process2);
    \draw [arrow] (process2) -- (process3);
    \draw [arrow] (process3) -- (process4);
    \draw [arrow] (process4) -- (process5);
    \draw [arrow] (process5) -- (decision);

    \draw [arrow] (decision) -- node[anchor=north] {No} +( -5cm, -0cm) |- (process2);
    \draw [arrow] (decision) -- node[anchor=west] {Yes} (stop);


    \end{tikzpicture}
    \caption{Flowchart of the Bayesian Optimization Loop. The process iteratively updates the surrogate model and selects new hyperparameters to evaluate until convergence criteria are met.}
    \label{fig:bayes_opt_flowchart}
\end{figure}
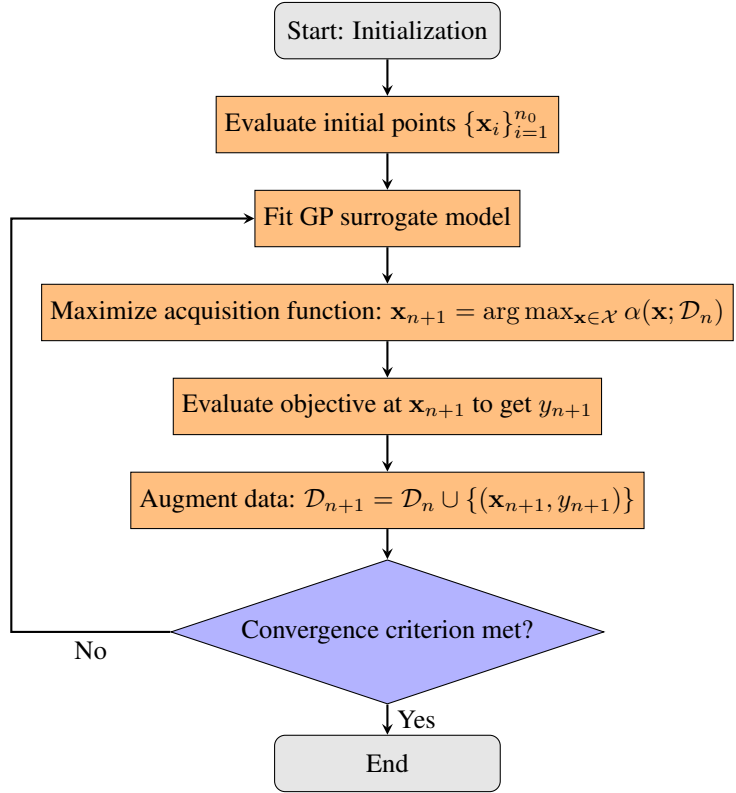

As shown in \figref{fig:bayes_opt_flowchart}, the Bayesian optimization algorithm proceeds iteratively through initialization, surrogate model updates, and acquisition function optimization.

The Bayesian optimization algorithm begins with the initialization step, where the objective function is evaluated at an initial set of hyperparameters $\{ \mathbf{x}_i \}_{i=1}^{n_0}$, often chosen via a space-filling design like Latin hypercube sampling. Next, a Gaussian Process (GP) surrogate model is fitted to the observed data $\mathcal{D}_n$, capturing our current understanding of the objective function.

Using this surrogate model, the algorithm maximizes the acquisition function to find the next promising set of hyperparameters:

\begin{equation}
\mathbf{x}_{n+1} = \arg\max_{\mathbf{x} \in \mathcal{X}} \alpha(\mathbf{x}; \mathcal{D}_n).
\end{equation}

The objective function is then evaluated at $\mathbf{x}_{n+1}$ to obtain $y_{n+1}$, and the data is augmented:

\begin{equation}
\mathcal{D}_{n+1} = \mathcal{D}_n \cup \{ (\mathbf{x}_{n+1}, y_{n+1}) \}.
\end{equation}

The algorithm checks if the convergence criterion is met (e.g., maximum iterations reached or negligible improvement observed). If not, the process loops back to updating the surrogate model with the new data, as depicted in \figref{fig:bayes_opt_flowchart}. This iterative process continues until convergence, resulting in the optimal set of hyperparameters.

\subsubsection{Parameter Space}

We optimized 13 hyperparameters of the image processing algorithm, each within specified bounds informed by prior knowledge as shown in \tabref{table:hyperparameters}. The ranges were chosen to cover a wide search space while ensuring that the parameters were physically meaningful and relevant to the task of detecting crystalline regions in HRTEM images.

\begin{table}[t!]
    \caption{Hyperparameters and their ranges used in Bayesian optimization.}
    \label{table:hyperparameters}
    \setlength{\extrarowheight}{2pt}
    \centering
    \begin{tabular}{||c|c|c||} 
    \hline
    \textbf{Hyperparameter} & \textbf{Range} & \textbf{Type} \\
    \hline\hline
    $\texttt{blur\_iteration}$          & [5, 20]          & Integer \\
    \hline
    $\texttt{Blur\_kernel\_propCons}$   & [0.1, 0.5]       & Real \\
    \hline
    $\texttt{closing\_k\_size}$         & [1, 20]          & Integer \\
    \hline
    $\texttt{opening\_k\_size}$         & [1, 20]          & Integer \\
    \hline
    $\texttt{pixThresh\_propCons}$      & [0.0, 1.0]       & Real \\
    \hline
    $\texttt{ellipse\_len\_propCons}$   & [0.5, 5.0]       & Real \\
    \hline
    $\texttt{ellipse\_aspect\_ratio}$   & [2.0, 7.0]       & Real \\
    \hline
    $\texttt{thresh\_dist\_propCons}$   & [1.0, 5.0]       & Real \\
    \hline
    $\texttt{thresh\_theta}$            & [5.0, 15.0]      & Real \\
    \hline
    $\texttt{cluster\_size}$            & [1, 10]          & Integer \\
    \hline
    $\texttt{dspace\_bandpass}$         & [0.1, 0.5]       & Real \\
    \hline
    $\texttt{powSpec\_peak\_thresh}$    & [1.0, 1.5]       & Real \\
    \hline
    $\texttt{thresh\_area\_factor}$     & [1.0, 5.0]       & Real \\
    \hline
    \end{tabular}
\end{table}

\subsubsection{Objective Function Evaluation}

\begin{figure}[t!]
    \centering
    \begin{tabular}{>{\centering\arraybackslash}m{0.29\textwidth} 
                    >{\centering\arraybackslash}m{0.29\textwidth} 
                    >{\centering\arraybackslash}m{0.29\textwidth}}
        \toprule
        \textbf{Input Image} & \textbf{Ground Truth Annotation} & \textbf{Grount Truth Mask} \\
        \midrule
        \includegraphics[width=\linewidth]{figs/train_input_FoilHole_21830243_Data_21829764_21829765_20200122_1102.jpg} &
        \includegraphics[width=\linewidth]{figs/train_GT_FoilHole_21830243_Data_21829764_21829765_20200122_1102.jpg} &
        \includegraphics[width=\linewidth]{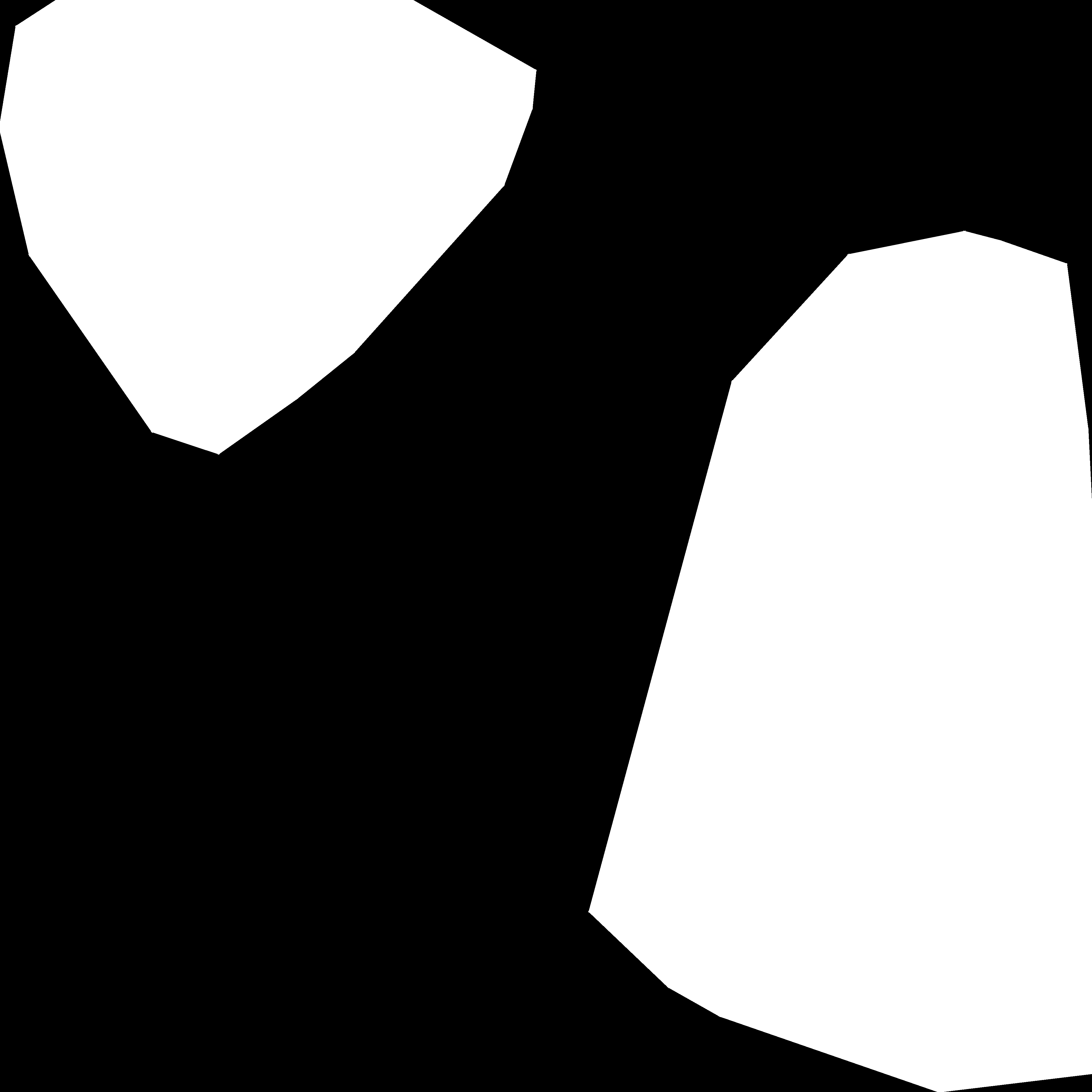} \\
        \midrule
        \includegraphics[width=\linewidth]{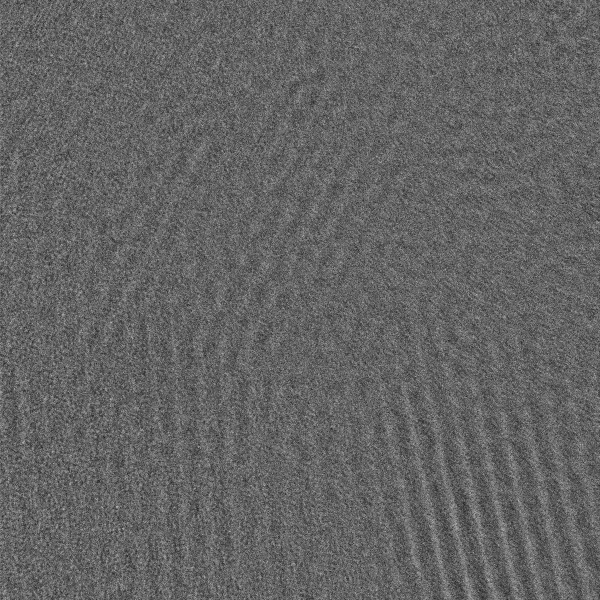} &
        \includegraphics[width=\linewidth]{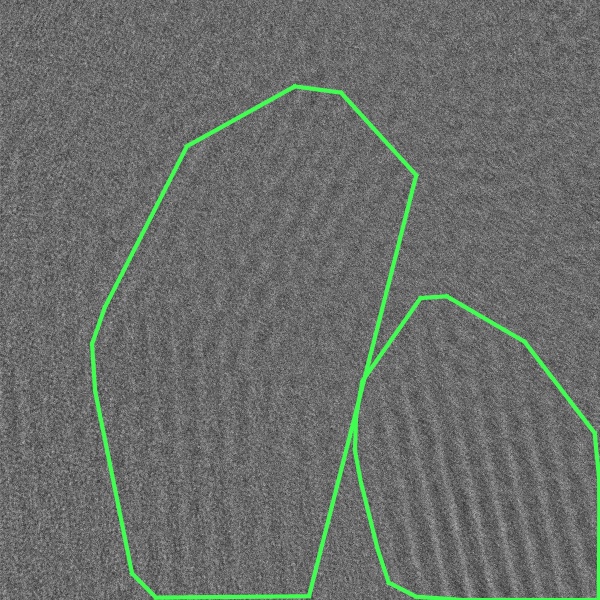} &
        \includegraphics[width=\linewidth]{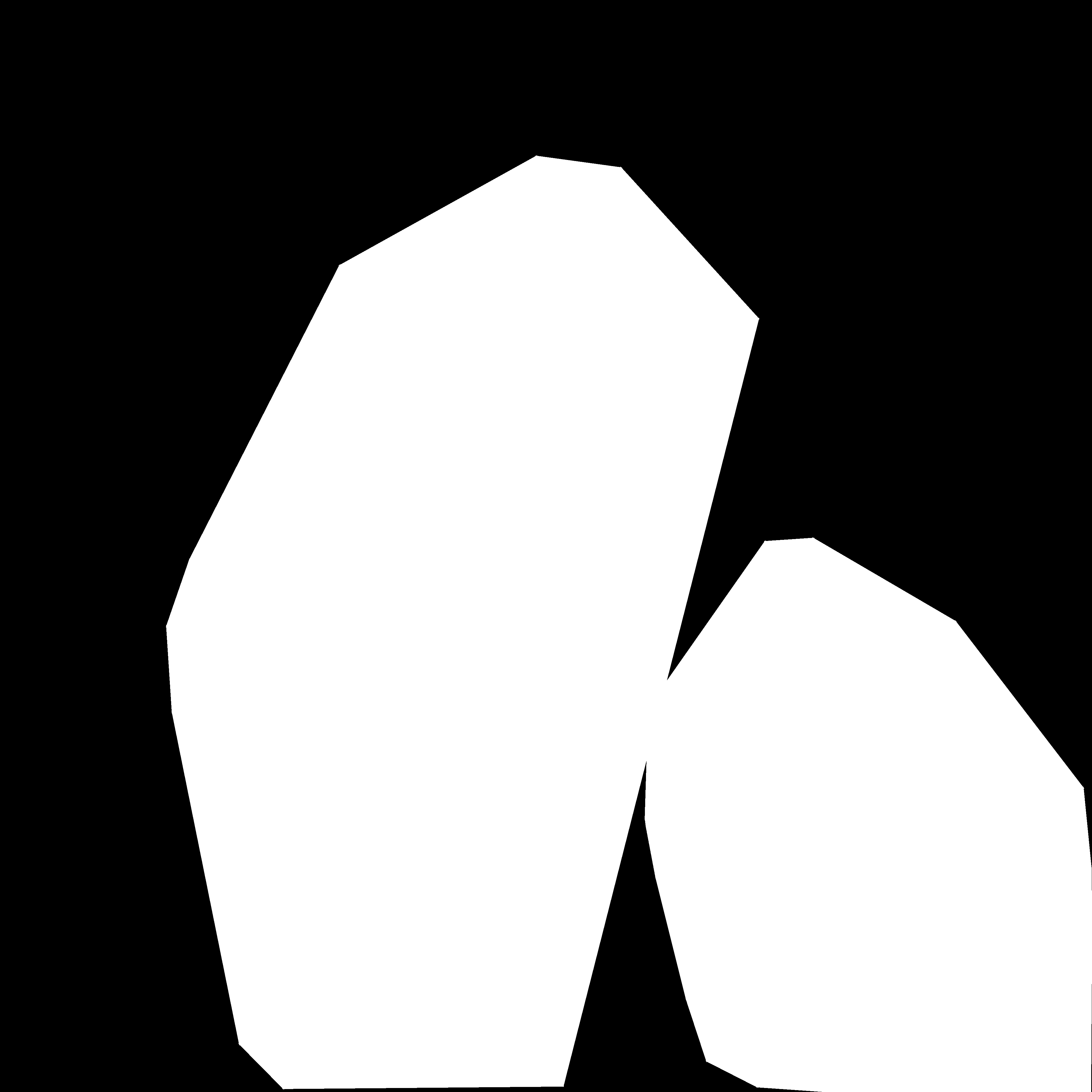} \\
        \midrule
        \includegraphics[width=\linewidth]{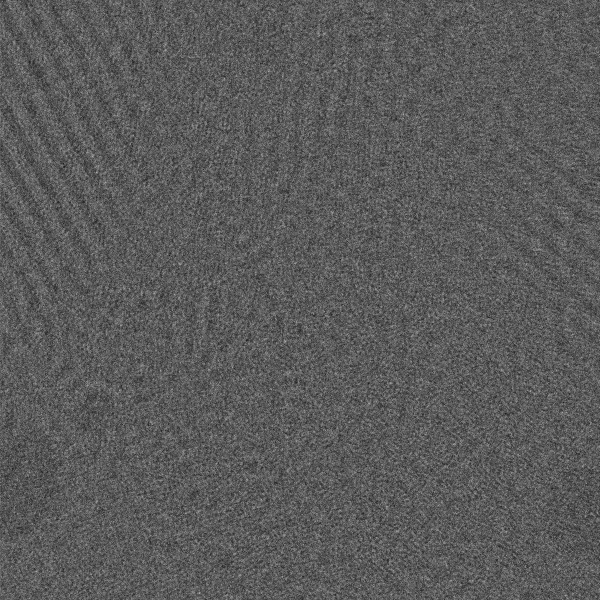} &
        \includegraphics[width=\linewidth]{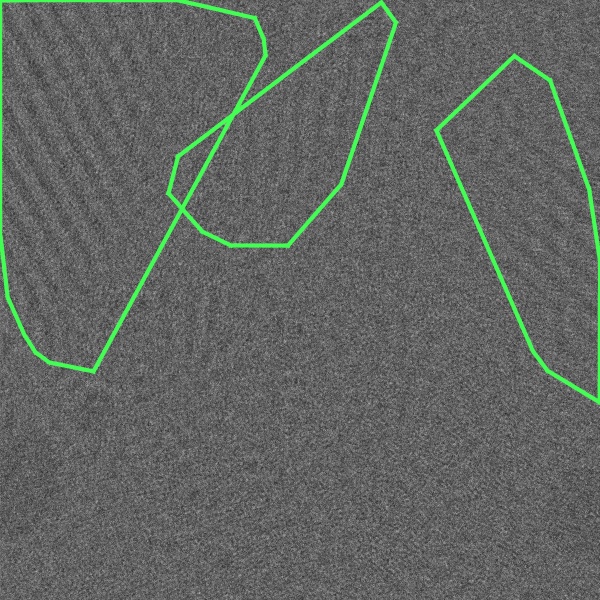} &
        \includegraphics[width=\linewidth]{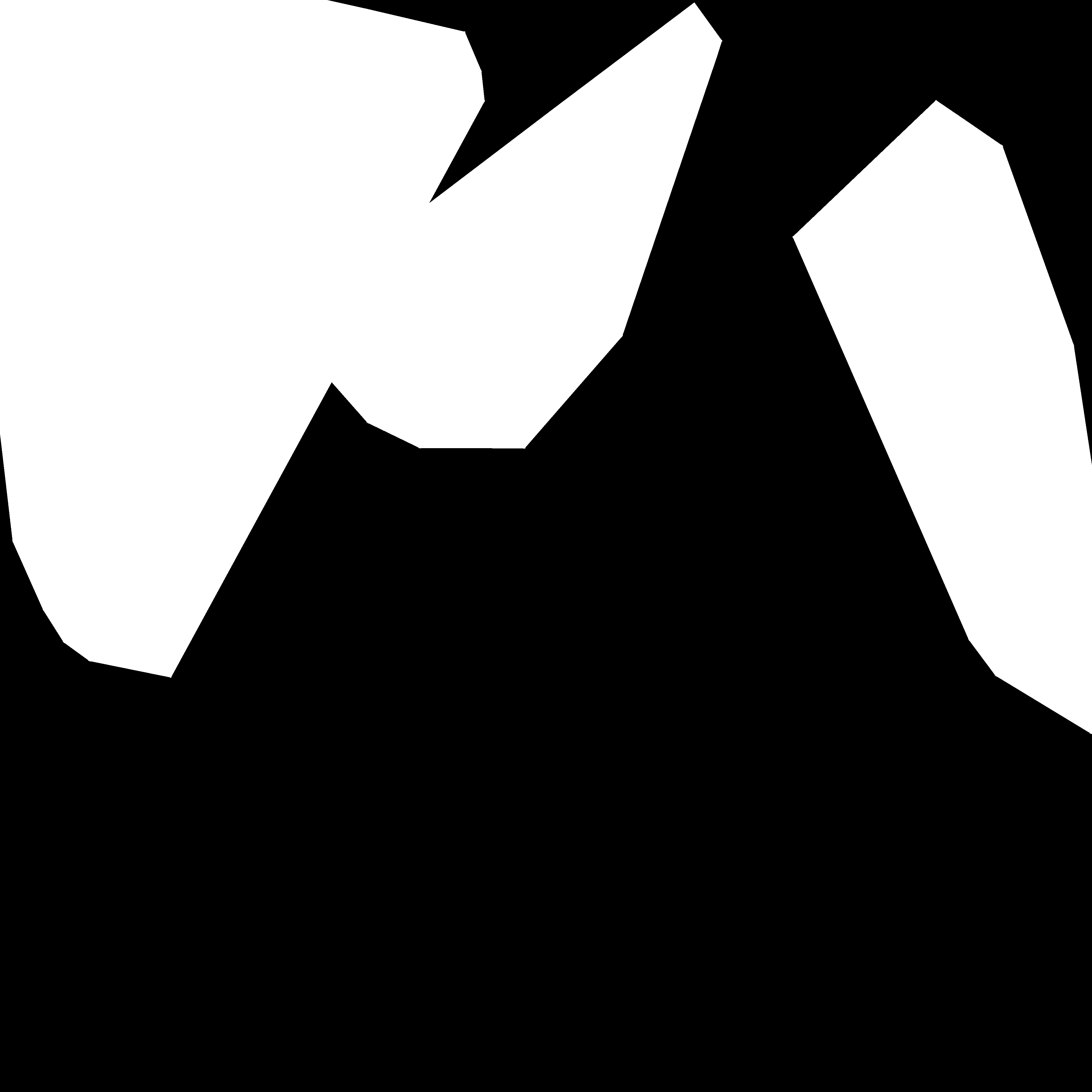} \\
        \bottomrule
    \end{tabular}
    \caption{Bayesian optimization training input i.e. input images, ground truth annotations, and ground truth masks. Each row represents a different sample from the dataset.}
    \label{fig:input_annotation_mask}
\end{figure}

\figref{fig:input_annotation_mask} shows example inputs for the Bayesian optimization training. We used a total of 13 annotated HRTEM images for training. The ground truth annotations are created using VGG annotator tool \citep{dutta2019vgg}. For each set of hyperparameters $\mathbf{x}$, the objective function $f(\mathbf{x})$ was evaluated as follows:

\begin{enumerate}
    \item Run the image processing algorithm with parameters $\mathbf{x}$ on the annotated HRTEM images.
    \item Generate binary masks of the detected crystalline regions.
    \item Compute the IoU between the detected masks and the ground truth masks:

    \begin{equation}
    \text{IoU}(\mathbf{x}) = \frac{|\text{Detected} \cap \text{Ground Truth}|}{|\text{Detected} \cup \text{Ground Truth}|},
    \end{equation}

    where $|\cdot|$ denotes the cardinality (number of pixels in this context).
    \item Set the objective function value as $f(\mathbf{x}) = -\text{IoU}(\mathbf{x})$.
\end{enumerate}

\subsubsection{Algorithm Execution}

Our implementation of the Gaussian Process-based Bayesian optimization algorithm for hyperparameter tuning is based on the \texttt{scikit-optimize} library \citep{head2021scikit}. The library provides a user-friendly interface through the \texttt{gp\_mini\-mize} function for Bayesian optimization and supports various acquisition functions, surrogate models, and optimization strategies. We used the \texttt{gp\_minimize} function with the following key parameters:

\begin{itemize}
    \item \textbf{Acquisition Function}: Expected Improvement (EI).
    \item \textbf{Number of Calls}: 200 total evaluations of the objective function.
    \item \textbf{Initial Points}: 10 random initial evaluations to seed the GP model.
    \item \textbf{Random State}: Seeded for reproducibility.
\end{itemize}

\subsection{Evaluation of Data Sufficiency}
\label{sec:data_sufficiency}

Determining the optimal amount of data to collect is crucial for experimentalists. Collecting too little data can compromise the reliability of results, while collecting excessive data may not yield additional insights and can waste valuable resources. Establishing a reliable stopping criterion for data collection ensures that resources are utilized efficiently without sacrificing statistical significance.

An effective data sufficiency metric should possess certain desirable features:

\begin{enumerate}
    \item \textbf{Sensitivity to Distribution Changes}: It should accurately reflect changes in the underlying data distribution as more data is collected.
    \item \textbf{Scale Interpretability}: The metric should provide a quantitative measure that is interpretable and can be related to practical thresholds for decision-making.
    \item \textbf{Applicability to Empirical Distributions}: It should be suitable for comparing empirical distributions derived from finite samples.
    \item \textbf{Metric Properties}: The measure should satisfy the properties of a mathematical metric, such as non-negativity, identity of indiscernibles, symmetry, and triangle inequality, to ensure consistent and meaningful comparisons.
\end{enumerate}

In this study, we introduce a stopping criterion based on the Wasserstein distance to assess data sufficiency. The Wasserstein distance, also known as the Earth Mover's Distance, quantifies the difference between two probability distributions by measuring the minimum "cost" of transforming one distribution into the other. It is particularly well-suited for our purposes because it satisfies all the desired features of a data sufficiency metric listed above.

\paragraph{Mathematical Definition of the Wasserstein Distance}

Let \( P \) and \( Q \) be two probability distributions on the real line with cumulative distribution functions (CDFs) \( F_P(x) \) and \( F_Q(x) \), respectively. The \( p \)-th order Wasserstein distance \( W_p(P, Q) \) between \( P \) and \( Q \) is defined as:

\begin{equation}
W_p(P, Q) = \left( \int_{-\infty}^{\infty} |F_P^{-1}(u) - F_Q^{-1}(u)|^p \, du \right)^{1/p}
\end{equation}

where \( F_P^{-1}(u) \) and \( F_Q^{-1}(u) \) are the quantile functions (inverse CDFs) of \( P \) and \( Q \), and \( p \geq 1 \). For \( p = 1 \), the first-order Wasserstein distance simplifies to:

\begin{equation}
W_1(P, Q) = \int_{0}^{1} |F_P^{-1}(u) - F_Q^{-1}(u)| \, du
\end{equation}

Alternatively, for discrete empirical distributions derived from finite samples, the first-order Wasserstein distance between two sets of observations \( \{ x_i \}_{i=1}^n \) and \( \{ y_j \}_{j=1}^m \) can be computed by sorting the observations and calculating the average absolute difference between the sorted values:

\begin{equation}
W_1(P_n, Q_m) = \frac{1}{N} \sum_{k=1}^{N} | x_{(k)} - y_{(k)} |
\end{equation}

where \( N = \min(n, m) \), and \( x_{(k)} \), \( y_{(k)} \) are the ordered statistics (sorted data).

The Wasserstein distance's ability to capture differences between distributions, even when they have overlapping support, makes it ideal for assessing the convergence of empirical data distributions as more data is collected. By monitoring the Wasserstein distance between successive data samples, we can determine when the distributions have converged, indicating that the data collection process has reached a point of diminishing returns.

\section*{Data Availability}
The dataset used in this work will be made available on request.

\section*{Code Availability}
\sloppy
 The software developed for this paper is available at \url{https://bitbucket.org/baskargroup/gratev2}.

\section*{Acknowledgements}
Funding from the National Science Foundation under Award DMREF 2323716 and the Office of Naval Research under Awards N00014-19-1-2453, and N00014-23-1-2001 are gratefully acknowledged. 

\section*{Conflict of Interest Statement}
The authors have no competing interests to declare that are relevant to the content of this article.

\clearpage
{\small
\putbib
}
\end{bibunit}

\clearpage
\begin{bibunit}
\appendix
\setcounter{page}{1}
\renewcommand\thefigure{A.\arabic{figure}}    
\setcounter{figure}{0}

\begin{center}
\openup 0.5em
{\usefont{OT1}{phv}{b}{n}\selectfont\Large{Appendix}}
\end{center}

\section{Optimum Parameters}
\label{appendix:OptimumParameters}

The Bayesian optimization based optimum parameters and manually selected parameters of \GRATEVtwo{} for our dataset are given in \tabref{table:parameters}.

\begin{table}[ht!]
    \centering
    \small
    \caption{Comparison of Manually selected and Bayesian Optimized Parameters}
    \label{table:parameters}
    \begin{tabular}{||c|c|c|c||}
        \hline
        \textbf{Parameter} & \textbf{Description} & \textbf{Manually} & \textbf{Bayesian} \\
            &  & \textbf{Selected Value} & \textbf{Optimized Value} \\
        \hline\hline
        $\texttt{dspace\_nm}$               & d-Spacing in nm & 1.9    & 1.9 \\
        \hline
        $\texttt{pix\_2\_nm}$               & pixels per nm & 78.5    & 78.5 \\
        \hline
        $\texttt{blur\_iteration}$          & Blurring Iterations & 15    & 20 \\
        \hline
        $\texttt{Blur\_kernel\_propCons}$   & Kernel Size Blurring* & 0.15  & 0.12 \\
        \hline
        $\texttt{closing\_k\_size}$         & Kernel Size Closing & 15    & 2 \\
        \hline
        $\texttt{opening\_k\_size}$         & Kernel Size Opening & 17    & 2 \\
        \hline
        $\texttt{pixThresh\_propCons}$      & Threshold Pixel Length* & 0.63  & 0.74 \\
        \hline
        $\texttt{ellipse\_len\_propCons}$   & Uniform Breaking Length* & 1.50  & 4.03 \\
        \hline
        $\texttt{ellipse\_aspect\_ratio}$   & Threshold Ellipse Aspect Ratio & 5.00  & 4.38 \\
        \hline
        $\texttt{thresh\_dist\_propCons}$   & Adjacency Distance* & 2.00  & 1.36 \\
        \hline
        $\texttt{thresh\_theta}$            & Adjacency Angle (degrees) & 10.00 & 13.96 \\
        \hline
        $\texttt{cluster\_size}$            & Threshold Cluster Size & 7     & 9 \\
        \hline
        $\texttt{dspace\_bandpass}$         & Band Pass Filter Size & 0.20  & 0.44 \\
        \hline
        $\texttt{powSpec\_peak\_thresh}$    & Power Spectrum Threshold & 1.15  & 1.00 \\
        \hline
        $\texttt{thresh\_area\_factor}$     & Area Threshold Factor & 4.00  & 2.79 \\
        \hline
    \end{tabular}
    \vspace{0.5em}
    \begin{itemize}
        \item $*$ Proportionality constant to $d-Spacing$.
        \item $\texttt{dspace\_nm}$ and $\texttt{pix\_2\_nm}$ are user inputs and not optimized.
    \end{itemize}
\end{table}

\section{T-statistic for Data Sufficiency}
\label{appendix:Tstatistic}

To evaluate the statistical significance of the improvement, we performed a paired t-test on the IoU scores from the two sets of parameters. The null hypothesis ($H_0$) is that there is no difference in the mean IoU scores between the manual and Bayesian-optimized parameters. Let $d_i$ be the difference in IoU scores for each image, defined as $d_i = \text{IoU}_{\text{Bayesian}, i} - \text{IoU}_{\text{Manual}, i}$. 

\begin{equation}
    \bar{d} = \frac{1}{n} \sum_{i=1}^{n} d_i = 0.1413
\end{equation}

\begin{equation}
    s_d = \sqrt{\frac{1}{n-1} \sum_{i=1}^{n} (d_i - \bar{d})^2} = 0.0489
\end{equation}

The t-statistic is computed as:

\begin{equation}
t = \frac{\bar{d}}{s_d / \sqrt{n}} = \frac{0.1413}{0.0489 / \sqrt{6}} = \frac{0.1413}{0.01997} = 7.074.
\end{equation}

With $n - 1 = 5$ degrees of freedom, the critical t-value at a significance level of $\alpha = 0.05$ (two-tailed) is approximately 2.571. Since $t = 7.074 > 2.571$, we reject the null hypothesis and conclude that the improvement in IoU scores using Bayesian optimization is statistically significant.

\clearpage

\section{Additional Validation Results}
\label{appendix:AdditionalValidationResults}

\begin{figure}[h!]
    \centering
    \begin{tabular}{>{\centering\arraybackslash}m{0.28\textwidth} 
                    >{\centering\arraybackslash}m{0.28\textwidth} 
                    >{\centering\arraybackslash}m{0.28\textwidth}}
        \toprule
        \textbf{Ground truth} & \textbf{Results using manually selected parameters} & \textbf{Results using Bayesian-optimized parameters} \\
        \midrule
        \includegraphics[width=\linewidth]{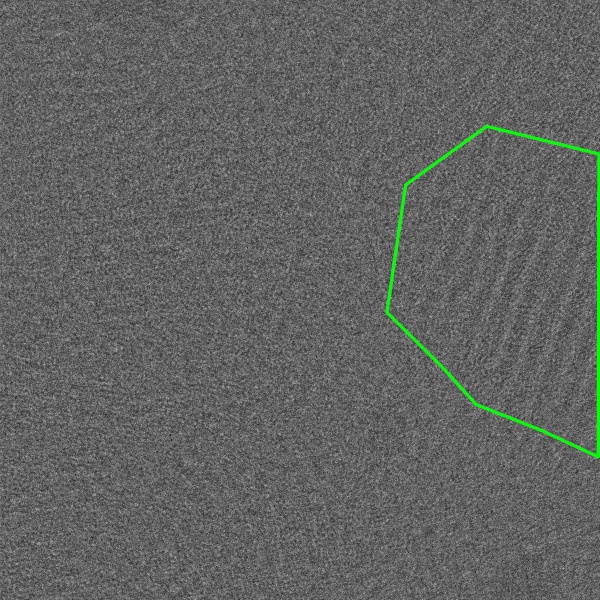} &
        \includegraphics[width=\linewidth]{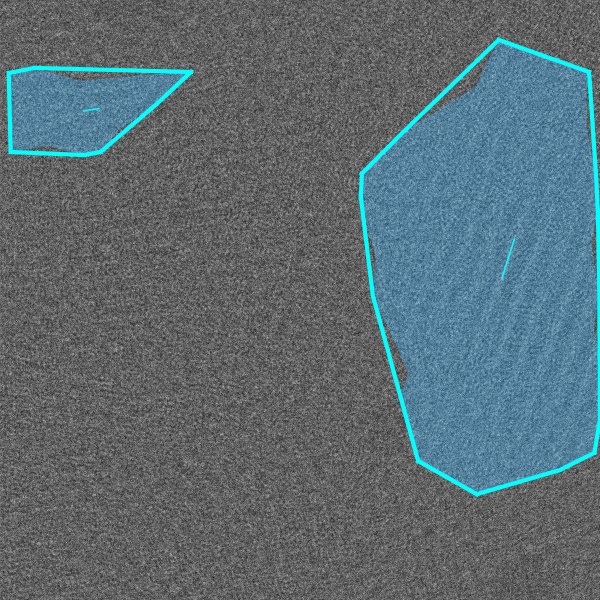} &
        \includegraphics[width=\linewidth]{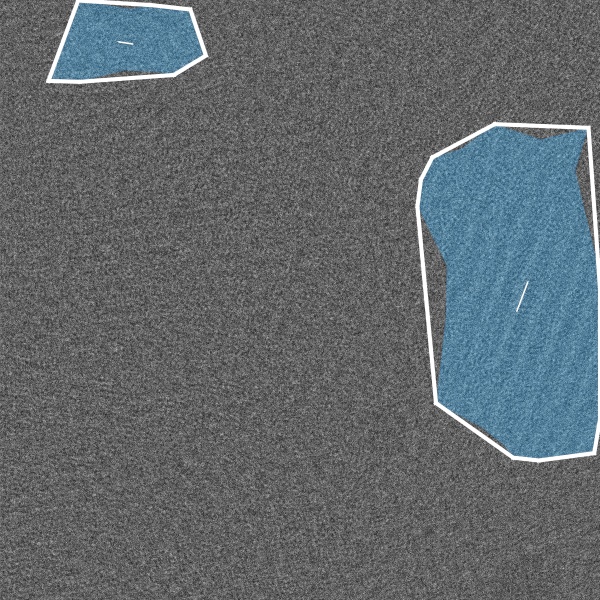} \\
        \midrule
        \includegraphics[width=\linewidth]{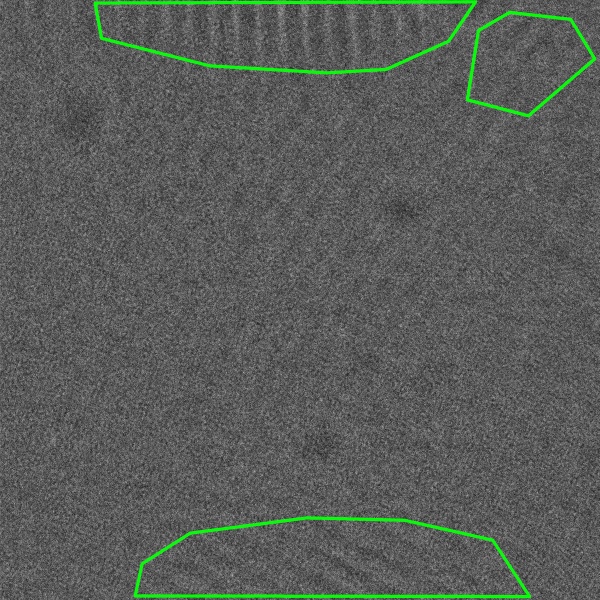} &
        \includegraphics[width=\linewidth]{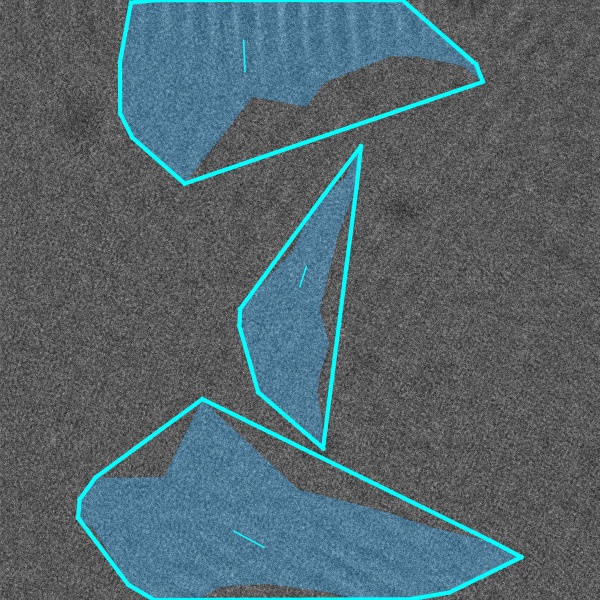} &
        \includegraphics[width=\linewidth]{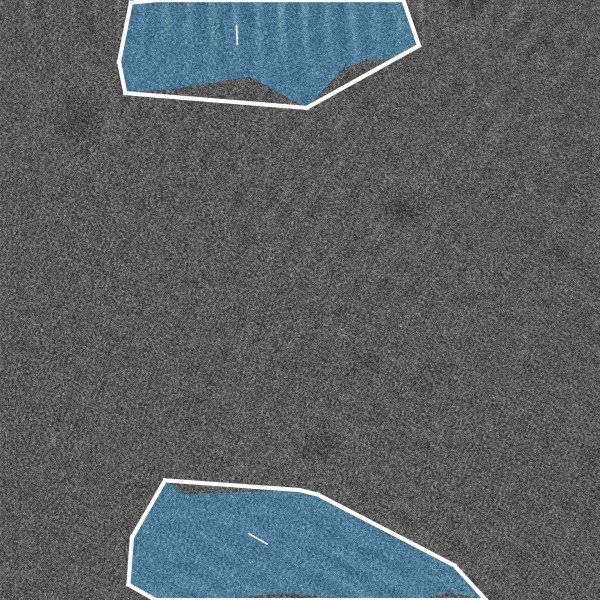} \\
        \midrule
        \includegraphics[width=\linewidth]{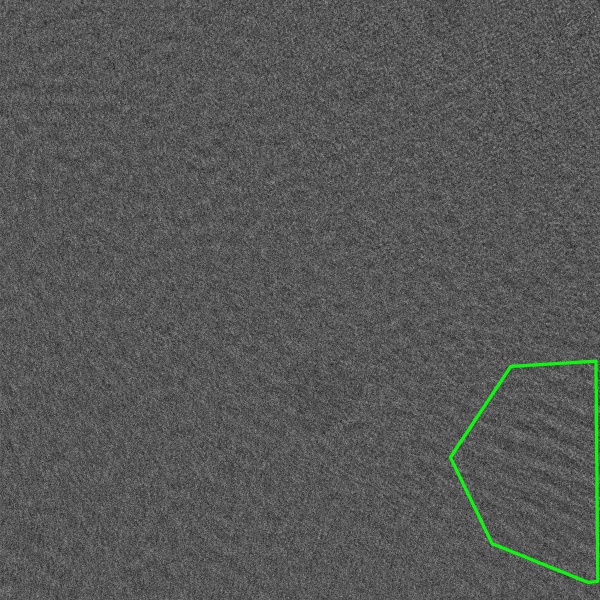} &
        \includegraphics[width=\linewidth]{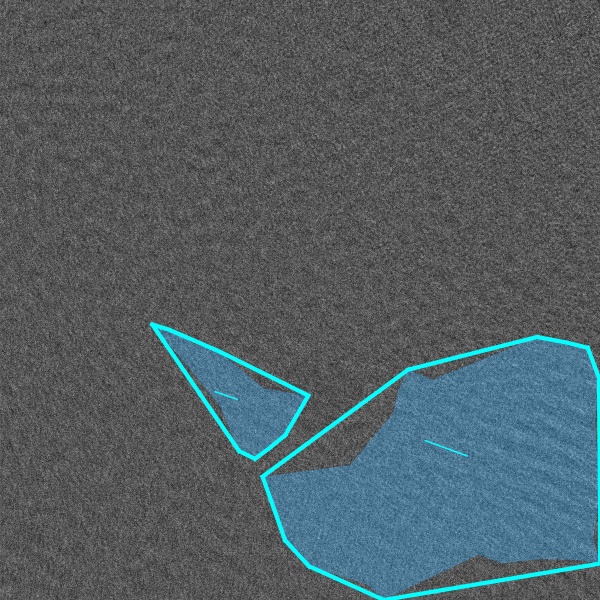} &
        \includegraphics[width=\linewidth]{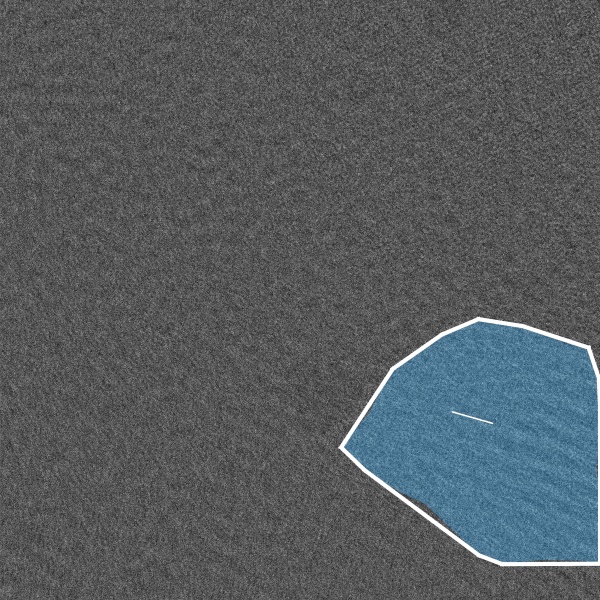} \\
        \bottomrule
    \end{tabular}
    \caption{Additional comparison of ground truth, manually selected parameters, and Bayesian-optimized parameters across different images.}
    \label{fig:additional_validation_results}
\end{figure}

\end{bibunit}

\end{document}